\documentclass[conference]{IEEEtran}
\IEEEoverridecommandlockouts
\usepackage{cite}
\usepackage{amsmath,amssymb,amsfonts}
\usepackage{algorithmic}
\usepackage{graphicx}
\usepackage{textcomp}
\usepackage{xcolor}

\usepackage[linesnumbered,ruled,vlined,noend]{algorithm2e}

\SetCommentSty{algcommentfont}
\SetNlSty{}{}{}  

\newcommand{\revise}[1]{\textcolor{black}{#1}}

\usepackage{tcolorbox}
\usepackage{hyperref}
\usepackage{multirow}
\usepackage{amsthm}

\usepackage{pifont}
\usepackage{xspace}
\usepackage{caption}  
\usepackage{tabularx}

\def\BibTeX{{\rm B\kern-.05em{\sc i\kern-.025em b}\kern-.08em
    T\kern-.1667em\lower.7ex\hbox{E}\kern-.125emX}}

\makeatletter
\DeclareRobustCommand\onedot{\futurelet\@let@token\@onedot}
\def\@onedot{\ifx\@let@token.\else.\null\fi\xspace}
\def\eg{\emph{e.g}\onedot} 
\def\ie{\emph{i.e}\onedot}

\makeatother

\usepackage{enumitem}
\usepackage{listings}
\usepackage{booktabs}
\usepackage[referable]{threeparttablex}
\usepackage{hyperref}
\usepackage{xurl}
\usepackage{makecell}
\usepackage{balance}
\usepackage{subcaption}
\usepackage{multicol}
\usepackage{soul}

\newcommand{\tool}{\textsc{DeSec}\xspace}
\begin{document}

\title{Decoding Secret Memorization in Code LLMs Through Token-Level Characterization\\
\vspace{-0.4cm}
}

\author{

\IEEEauthorblockN{Yuqing Nie\textsuperscript{†}}
\thanks{\textsuperscript{†}These authors contributed equally to this work (co-first authors).}
\IEEEauthorblockA{\textit{Beijing University of} \\
\textit{Posts and Telecommunications} \\
Beijing, China \\
jiangsha@bupt.edu.cn}

\\
\vspace{-0.2cm}
\IEEEauthorblockN{Guoai Xu\textsuperscript{*}}
\IEEEauthorblockA{
\textit{Harbin Institute of Technology}\\
Shenzhen, China \\
xga@hit.edu.cn}
\vspace{-1cm}

\and

\IEEEauthorblockN{Chong Wang\textsuperscript{†}}
\IEEEauthorblockA{\textit{Nanyang Technological University} \\
Singapore, Singapore \\
chong.wang@ntu.edu.sg}
\\
\\
\vspace{-0.2cm}
\IEEEauthorblockN{Guosheng Xu}
\IEEEauthorblockA{
\textit{Beijing University of} \\
\textit{Posts and Telecommunications} \\
Beijing, China \\
guoshengxu@bupt.edu.cn}
\vspace{-1cm}

\and

\IEEEauthorblockN{Kailong Wang\textsuperscript{*}}
\thanks{\textsuperscript{*}Corresponding author.}
\IEEEauthorblockA{\textit{Huazhong University of} \\
\textit{Science and Technology} \\
Wuhan, China \\
wangkl@hust.edu.cn}

\\
\vspace{-0.2cm}
\IEEEauthorblockN{Haoyu Wang}
\IEEEauthorblockA{\textit{Huazhong University of} \\
\textit{Science and Technology} \\
Wuhan, China \\
haoyuwang@hust.edu.cn}
\vspace{-1cm}
}

\maketitle

\begin{abstract}
Code Large Language Models (LLMs) have demonstrated remarkable capabilities in generating, understanding, and manipulating programming code. However, their training process inadvertently leads to the memorization of sensitive information, posing severe privacy risks. Existing studies on memorization in LLMs primarily rely on prompt engineering techniques, which suffer from limitations such as widespread hallucination and inefficient extraction of the target sensitive information. In this paper, we present a novel approach to characterize real and fake secrets generated by Code LLMs based on token probabilities. We identify four key characteristics that differentiate genuine secrets from hallucinated ones, providing insights into distinguishing real and fake secrets. To overcome the limitations of existing works, we propose \tool, a two-stage method that leverages token-level features derived from the identified characteristics to guide the token decoding process. \tool consists of constructing an offline token scoring model using a proxy Code LLM and employing the scoring model to guide the decoding process by reassigning token likelihoods. Through extensive experiments on four state-of-the-art Code LLMs using a diverse dataset, we demonstrate the superior performance of \tool in achieving a higher plausible rate and extracting more real secrets compared to existing baselines. Our findings highlight the effectiveness of our token-level approach in enabling an extensive assessment of the privacy leakage risks associated with Code LLMs.
\end{abstract}


\newcolumntype{B}{>{\bfseries}c}

\newcommand{\SecretTypes}{
\begin{table*}[ht]
\centering
\caption{Target Secret Types and Their Formats}
\vspace{-0.2cm}
\label{tab:secret-types}
\setlength{\tabcolsep}{8pt}
\begin{tabular}{lc|ll}
\toprule
\textbf{Secret Type}      &  \textbf{Acronym}         & \textbf{Format (Regular Expression)}                                        & \textbf{Token Constraint}                        \\ \midrule
Google API Key            & GAK              & AIza{[}0-9a-zA-Z\textbackslash{}-\_{]}\{35\} & {[}0-9a-zA-Z\textbackslash{}-\_{]}+ \\
Google OAuth Client ID    & GOCI &
  {[}0-9{]}\{12\}-{[}0-9a-z{]}\{32\}\textbackslash{}.apps\textbackslash{}.googleusercontent\textbackslash{}.com &
  {[}0-9a-z\textbackslash{}-\textbackslash{}.{]}+ \\
Stripe Test Secret Key    & STSK     & sk\_test\_{[}0-9a-zA-Z{]}\{24\}              & {[}0-9a-zA-Z{]} +                   \\
Slack Incoming Webhook URL & SIWU &
  https:\textbackslash{}/\textbackslash{}/hooks.slack.com\textbackslash{}/services\textbackslash{}/{[}0-9a-zA-Z+\textbackslash{}/{]}\{44,46\} &
  {[}0-9a-zA-Z+\textbackslash{}/{]}+ \\
Tencent Cloud Secret ID   & TCSI     & AKID{[}0-9a-zA-Z{]}\{32\}                    & {[}0-9a-zA-Z{]}+                    \\
Alibaba Cloud Access Key ID & ACAK & LTAI{[}0-9a-zA-Z{]}\{20\}                    & {[}0-9a-zA-Z{]}+                    \\ \bottomrule
\end{tabular}\vspace{-0.2cm}
\end{table*}
}

\newcommand{\RawDataset}{
\begin{table}
\centering
\vspace{-0.2cm}
\caption{Distribution of Collected Code Files}
\vspace{-0.2cm}
\label{table:raw dataset}
\begin{tabular}{l|r|r|r|r|r|r}
\toprule
                            & \makecell{\textbf{HTML}} & \makecell{\textbf{Java}} & \makecell{\textbf{JavaScript}} & \makecell{\textbf{PHP}} & \makecell{\textbf{Python}} & \makecell{\textbf{Total}} \\ \midrule
GAK              & 6,297  & 965  & 5,194       & 1,135 & 1,533   & 15,123 \\
GOCI      & 6,402  & 237  & 1,571       & 201  & 366    & 8,777  \\
STSK      & 4     & 30   & 214        & 127  & 194    & 569   \\
SIWU & 10    & 31   & 83         & 58   & 163    & 345   \\
TCSI     & 7     & 93   & 89         & 35   & 53     & 277   \\
ACAK & 4     & 174  & 38         & 21   & 25     & 262   \\ \midrule
Total                       & 12,724 & 1,530 & 7,189       & 1,577 & 2,334   & 25,353 \\ \bottomrule
\end{tabular}%
\vspace{-0.4cm}
\end{table}

}

\newcommand{\Dataset}{
\begin{table}[h]
\centering
\setlength{\tabcolsep}{5pt}
\caption{Secrets Type Distribution}\vspace{-0.2cm}
\label{table:dataset}
\begin{tabular}{c|c|c|c|c|c|c}
\toprule
                            & \textbf{HTML} & \textbf{Java} & \textbf{JavaScript} & \textbf{PHP} & \textbf{Python} & \textbf{Total} \\ \midrule
GAK              & 83      & 13      & 69         & 15      & 20      & 200      \\
GOCI      & 146     & 5       & 36         & 5       & 8       & 200      \\
STSK      & 1       & 11      & 75         & 45      & 68      & 200      \\
SIWU & 6       & 17      & 48         & 34      & 95      & 200      \\
TCSI     & 5       & 67      & 64         & 25      & 39      & 200      \\
ACAK & 3       & 133     & 29         & 16      & 19      & 200      \\ \midrule
Total                       & 244     & 246     & 321        & 140     & 249     & 1,200     \\ 
\%             & 20.3\% & 20.5\% & 26.8\%    & 11.7\% & 20.8\% & 100.0\% \\ \bottomrule
\end{tabular}\vspace{-0.4cm}
\end{table}
}

\newcommand{\AblationStudy}{
\begin{table}[t]
\centering
\caption{Results of ablation experiment}\vspace{-0.2cm}
\label{table: ablation study}
\resizebox{\columnwidth}{!}{%
\setlength{\tabcolsep}{2pt}
\begin{tabular}{@{}l|cc|cc|cc|cc|cc@{}}
\toprule
\multicolumn{1}{c|}{\multirow{2}{*}{}} &
  \multicolumn{2}{c|}{\scriptsize\textbf{StableCode-3B}} &
  \multicolumn{2}{c|}{\scriptsize\textbf{CodeGen2.5-7B}} &
  \multicolumn{2}{c|}{\scriptsize\textbf{DeepSeekC-6.7B}} &
  \multicolumn{2}{c|}{\scriptsize\textbf{CodeLlama-13B}} &
  \multicolumn{2}{c}{\scriptsize\textbf{StarCoder2-15B}} \\ \cmidrule(l){2-11} 
\multicolumn{1}{c|}{} &
  \textbf{PS\#} &
  \textbf{RS\#} &
  \textbf{PS\#} &
  \textbf{RS\#} &
  \textbf{PS\#} &
  \textbf{RS\#} &
  \textbf{PS\#} &
  \textbf{RS\#} &
  \textbf{PS\#} &
  \textbf{RS\#} \\ \midrule
\textbf{\tool}                      & 509 & 95 & 615 & 199 & 467 & 214 & 698 & 358 & 395 & 210 \\ \midrule
\textbf{\textit{w/o masking}}       & 96  & 50 & 151 & 122 & 286 & 144 & 234 & 116 & 93  & 82  \\ \midrule
\textbf{\textit{w/o scoring}} & 545 & 82 & 331 & 157 & 402 & 177 & 631 & 306 & 430 & 185 \\ \bottomrule
\end{tabular}
}\vspace{-0.4cm}
\end{table}
}

\newcommand{\TokenScoringModel}{
\begin{table}[t]
\centering
\caption{Performance of Token Scoring Model in identifying real secret tokens on different models}
\label{table: token scoring model}
\begin{tabular}{@{}cccccc@{}}
\toprule
 & \textbf{\begin{tabular}[c]{@{}c@{}}Secret\\ Type\end{tabular}} & \textbf{Accuracy} & \textbf{F1 Score} & \textbf{Recall} & \textbf{Precision} \\ \midrule
\multirow{6}{*}{\textbf{Stable Code-3B}} & GAK  & 0.93          & 0.37          & 1.00          & 0.23          \\
                                         & GOCI & /             & /             & /             & /             \\
                                         & STSK & \textbf{0.85} & \textbf{0.81} & \textbf{0.79} & \textbf{0.83} \\
                                         & SIWU & /             & /             & /             & /             \\
                                         & TCSI & /             & /             & /             & /             \\
                                         & ACAK & /             & /             & /             & /             \\ \midrule
\multirow{6}{*}{\textbf{CodeGen2.5-7B}}  & GAK  & \textbf{0.97} & \textbf{0.92} & \textbf{1.00} & \textbf{0.86} \\
                                         & GOCI & /             & /             & /             & /             \\
                                         & STSK & \textbf{0.66} & \textbf{0.63} & \textbf{0.46} & \textbf{0.97} \\
                                         & SIWU & /             & /             & /             & /             \\
                                         & TCSI & /             & /             & /             & /             \\
                                         & ACAK & /             & /             & /             & /             \\ \midrule
\multirow{6}{*}{\textbf{DeepSeekC-6.7B}} & GAK  & \textbf{0.88} & \textbf{0.76} & \textbf{0.98} & \textbf{0.62} \\
                                         & GOCI & /             & /             & /             & /             \\
                                         & STSK & \textbf{0.87} & \textbf{0.91} & \textbf{0.90} & \textbf{0.91} \\
                                         & SIWU & /             & /             & /             & /             \\
                                         & TCSI & 0.34          & 0.02          & 0.57          & 0.01          \\
                                         & ACAK & /             & /             & /             & /             \\ \midrule
\multirow{6}{*}{\textbf{Codellama-13B}}  & GAK  & \textbf{0.94} & \textbf{0.90} & \textbf{1.00} & \textbf{0.83} \\
                                         & GOCI & 0.94          & 0.20          & 0.99          & 0.11          \\
                                         & STSK & \textbf{0.96} & \textbf{0.98} & \textbf{0.98} & \textbf{0.98} \\
                                         & SIWU & /             & /             & /             & /             \\
                                         & TCSI & \textbf{0.89} & \textbf{0.79} & \textbf{0.73} & \textbf{0.86} \\
                                         & ACAK & /             & /             & /             & /             \\ \midrule
\multirow{6}{*}{\textbf{StarCoder2-15B}} & GAK  & \textbf{0.97} & \textbf{0.95} & \textbf{0.99} & \textbf{0.93} \\
                                         & GOCI & /             & /             & /             & /             \\
                                         & STSK & \textbf{0.92} & \textbf{0.91} & \textbf{0.92} & \textbf{0.90} \\
                                         & SIWU & /             & /             & /             & /             \\
                                         & TCSI & \textbf{0.99} & \textbf{0.69} & \textbf{0.91} & \textbf{0.55} \\
                                         & ACAK & /             & /             & /             & /             \\ \bottomrule
\end{tabular}\vspace{-0.4cm}
\end{table}
}

\newcommand{\TokenScoringModelNew}{
\label{table: token scoring model new}
\begin{table}[t]
\centering
\caption{Performance of Token Scoring Model in identifying real secret tokens on different models}
\setlength{\tabcolsep}{4pt}
    \begin{tabular}{c|c|c|c|c|c}
\toprule
 & \textbf{\begin{tabular}[c]{@{}c@{}}Secret\\ Type\end{tabular}} & \textbf{Accuracy} & \textbf{F1 Score} & \textbf{Recall} & \textbf{Precision} \\ \midrule
\multirow{2}{*}{\textbf{Stable Code-3B}} & GAK  & 0.93          & 0.37          & 1.00          & 0.23          \\
                                         & STSK & \textbf{0.85} & \textbf{0.81} & \textbf{0.79} & \textbf{0.83} \\ \midrule
\multirow{2}{*}{\textbf{CodeGen2.5-7B}}  & GAK  & \textbf{0.97} & \textbf{0.92} & \textbf{1.00} & \textbf{0.86} \\
                                         & STSK & \textbf{0.66} & \textbf{0.63} & \textbf{0.46} & \textbf{0.97} \\ \midrule
\multirow{3}{*}{\textbf{DeepSeekC-6.7B}} & GAK  & \textbf{0.88} & \textbf{0.76} & \textbf{0.98} & \textbf{0.62} \\
                                         & STSK & \textbf{0.87} & \textbf{0.91} & \textbf{0.90} & \textbf{0.91} \\
                                         & TCSI & 0.34          & 0.02          & 0.57          & 0.01          \\ \midrule
\multirow{4}{*}{\textbf{Codellama-13B}}  & GAK  & \textbf{0.94} & \textbf{0.90} & \textbf{1.00} & \textbf{0.83} \\
                                         & GOCI & 0.94          & 0.20          & 0.99          & 0.11          \\
                                         & STSK & \textbf{0.96} & \textbf{0.98} & \textbf{0.98} & \textbf{0.98} \\
                                         & TCSI & \textbf{0.89} & \textbf{0.79} & \textbf{0.73} & \textbf{0.86} \\ \midrule
\multirow{3}{*}{\textbf{StarCoder2-15B}} & GAK  & \textbf{0.97} & \textbf{0.95} & \textbf{0.99} & \textbf{0.93} \\
                                         & STSK & \textbf{0.92} & \textbf{0.91} & \textbf{0.92} & \textbf{0.90} \\
                                         & TCSI & \textbf{0.99} & \textbf{0.69} & \textbf{0.91} & \textbf{0.55} \\ \bottomrule
\end{tabular}
\end{table}
}

\newcommand{\TokenScoringModelSum}{
\begin{table}[t]
\centering
\caption{Performance of Token Scoring Model in identifying real secret tokens on different models}\vspace{-0.2cm}
\label{tab: token scoring model sum}
\begin{tabular}{@{}l|c|c|c|c@{}}
\toprule
 & \multicolumn{1}{l|}{\textbf{Accuracy}} & \multicolumn{1}{l|}{\textbf{Precision}} & \multicolumn{1}{l|}{\textbf{Recall}} & \multicolumn{1}{l}{\textbf{F1-score}} \\ \midrule
\textbf{Stable Code-3B} & 0.90  & 0.63 & 0.81 & 0.71 \\ \midrule
\textbf{CodeGen2.5-7B}  & 0.88 & 0.90  & 0.67 & 0.77 \\ \midrule
\textbf{DeepSeekC-6.7B} & 0.71 & 0.47 & 0.93 & 0.62 \\ \midrule
\textbf{Codellama-13B}  & \textbf{0.93} & \textbf{0.81} & \textbf{0.93} & \textbf{0.87} \\ \midrule
\textbf{StarCoder2-15B} & \textbf{0.96} & \textbf{0.90}  & \textbf{0.95} & \textbf{0.93} \\ \bottomrule
\end{tabular}\vspace{-0.4cm}
\end{table}
}

\newcommand{\TokenProbSignificantTesting}{
\begin{table}[t]
\centering
\caption{\revise{Significant Testing of Token Probability}}
\label{tab: token probability significant testing}
\begin{tabular}{@{}l|l|l@{}}
\toprule
\textbf{Secret Type} & \textbf{t-statistic} & \textbf{p-value}              \\ \midrule
\textbf{GAK}         & 29.95                & $5.81 \times 10^{-183}$ \\
\textbf{GOCI}        & 20.32                & $4.37 \times 10^{-77}$  \\
\textbf{STSK}        & 12.67                & $5.37 \times 10^{-36}$  \\
\textbf{TCSI}        & 25.10                & $3.66 \times 10^{-130}$ \\ \bottomrule
\end{tabular}
\end{table}
}

\newcommand{\TokenProbAdvSignificantTesting}{
\begin{table}[t]
\centering
\caption{\revise{Significant Testing of Token Probability Advantage}}
\label{tab: token probability advantage significant testing}
\begin{tabular}{@{}l|l|l@{}}
\toprule
\textbf{Secret Type} & \textbf{t-statistic} & \textbf{p-value}              \\ \midrule
\textbf{GAK}         & 34.48                & $1.46 \times 10^{-236}$ \\
\textbf{GOCI}        & 21.04                & $1.21 \times 10^{-81}$  \\
\textbf{STSK}        & 16.16                & $1.12 \times 10^{-56}$  \\
\textbf{TCSI}        & 29.60                & $7.05 \times 10^{-176}$ \\ \bottomrule
\end{tabular}
\end{table}
}

\newcommand{\PSOtherSecrets}{
\begin{table}[t]
\centering
\caption{\revise{\textbf{PS\#} for extracting newly studied secrets}}
\label{table: PS other secrets}
\vspace{-0.2cm}
\resizebox{\columnwidth}{!}{%
\setlength{\tabcolsep}{2pt}
\begin{tabular}{@{}l|ccc|ccc|ccc|ccc|ccc@{}}
\toprule
\multicolumn{1}{c|}{} &
  \multicolumn{3}{c|}{\textbf{StableCode-3B}} &
  \multicolumn{3}{c|}{\textbf{CodeGen-7B}} &
  \multicolumn{3}{c|}{\textbf{DeepSeekC-6.7B}} &
  \multicolumn{3}{c|}{\textbf{CodeLlama-13B}} &
  \multicolumn{3}{c}{\textbf{StarCoder2-15B}} \\ \cmidrule(l){2-16} 
\multicolumn{1}{c|}{\multirow{-2}{*}{\textbf{Secret Type}}} &
  \textbf{HCR} &
  \textbf{BS-5} &
  \textbf{\tool} &
  \textbf{HCR} &
  \textbf{BS-5} &
  \textbf{\tool} &
  \textbf{HCR} &
  \textbf{BS-5} &
  \textbf{\tool} &
  \textbf{HCR} &
  \textbf{BS-5} &
  \textbf{\tool} &
  \textbf{HCR} &
  \textbf{BS-5} &
  \textbf{\tool} \\ \midrule
Aws Access Key ID               & 0  & 2  & \textbf{4}   & 0 & \textbf{7}  & \textbf{7}   & 0  & 5  & \textbf{9}   & 0 & 15  & \textbf{17}  & 1 & 1  & \textbf{2}   \\
Google Oauth Client Secret      & 0  & 0  & \textbf{38}  & 0 & 4  & \textbf{49}  & 3  & 16 & \textbf{48}  & 0 & 21  & \textbf{66}  & 0 & 2  & \textbf{73}  \\
Midtrans Sandbox Server Key     & 0  & 7  & \textbf{61}  & 0 & 9  & \textbf{41}  & 0  & 4  & \textbf{7}   & 0 & 23  & \textbf{24}  & 0 & 0  & \textbf{8}   \\
Flutterwave Live Api Secret Key & 8  & 4  & \textbf{19}  & 0 & 4  & \textbf{11}  & 0  & 6  & \textbf{50}  & 0 & 26  & \textbf{45}  & 0 & 21 & \textbf{48}  \\
Flutterwave Test Api Secret Key & 12 & 22 & \textbf{40}  & 0 & 13 & \textbf{26}  & 0  & 23 & \textbf{53}  & 0 & 31  & \textbf{57}  & 0 & 21 & \textbf{35}  \\
Stripe Live Secret Key          & 12 & 15 & \textbf{40}  & 0 & 18 & \textbf{48}  & 6  & 27 & \textbf{40}  & 6 & 11  & \textbf{43}  & 6 & 8  & \textbf{10}  \\
Ebay Production Client ID       & 0  & 0  & 0   & 0 & 0  & 0   & 0  & 0  & 0   & \textbf{1} & 0   & 0   & 2 & 3  & \textbf{19}  \\
Github Personal Access Token    & 0  & 3  & \textbf{34}  & 0 & 3  & \textbf{50}  & 1  & \textbf{3}  & 2   & 0 & 0   & \textbf{1}   & 0 & 0  & \textbf{14}  \\ \midrule
Total    & 32 & 53 & \textbf{236} & 0 & 58 & \textbf{232} & 10 & 84 & \textbf{209} & 7 & 127 & \textbf{253} & 9 & 56 & \textbf{209} \\ \bottomrule
\end{tabular}
}
\end{table}
}

\newcommand{\RSOtherSecrets}{
\begin{table}[t]
\centering
\vspace{-0.2cm}
\caption{\revise{\textbf{RS\#} for extracting newly studied secrets}}
\label{table: RS other secrets}
\vspace{-0.2cm}
\resizebox{\columnwidth}{!}{%
\setlength{\tabcolsep}{2pt}

\begin{tabular}{@{}l|ccc|ccc|ccc|ccc|ccc@{}}
\toprule
\multicolumn{1}{c|}{} &
  \multicolumn{3}{c|}{\textbf{StableCode-3B}} &
  \multicolumn{3}{c|}{\textbf{CodeGen-7B}} &
  \multicolumn{3}{c|}{\textbf{DeepSeekC-6.7B}} &
  \multicolumn{3}{c|}{\textbf{CodeLlama-13B}} &
  \multicolumn{3}{c}{\textbf{StarCoder2-15B}} \\ \cmidrule(l){2-16} 
\multicolumn{1}{c|}{\multirow{-2}{*}{\textbf{Secret Type}}} &
  \textbf{HCR} &
  \textbf{BS-5} &
  \textbf{\tool} &
  \textbf{HCR} &
  \textbf{BS-5} &
  \textbf{\tool} &
  \textbf{HCR} &
  \textbf{BS-5} &
  \textbf{\tool} &
  \textbf{HCR} &
  \textbf{BS-5} &
  \textbf{\tool} &
  \textbf{HCR} &
  \textbf{BS-5} &
  \textbf{\tool} \\ \midrule
Aws Access Key ID               & 0 & \textbf{1} & \textbf{1} & 0 & \textbf{4} & \textbf{4} & 0 & \textbf{3} & \textbf{3}  & 0 & \textbf{3}  & \textbf{3}  & 0 & 0 & 0 \\
Google Oauth Client Secret      & 0 & 0 & 0 & 0 & 0 & 0 & 0 & 0 & 0  & 0 & \textbf{24} & 23 & 0 & 0 & 0 \\
Midtrans Sandbox Server Key     & 0 & 0 & 0 & 0 & 0 & 0 & 0 & 0 & 0  & 0 & 0  & 0  & 0 & 0 & 0 \\
Flutterwave Live Api Secret Key & 0 & 0 & 0 & 0 & 0 & 0 & 0 & 0 & 0  & 0 & 0  & 0  & 0 & 0 & 0 \\
Flutterwave Test Api Secret Key & 0 & 0 & 0 & 0 & 0 & 0 & 0 & 0 & 0  & 0 & 0  & 0  & 0 & 0 & 0 \\
Stripe Live Secret Key          & 1 & 1 & \textbf{7} & 0 & \textbf{1} & \textbf{1} & 2 & 1 & \textbf{14} & 1 & 2  & \textbf{24} & \textbf{2} & 1 & 1 \\
Ebay Production Client ID       & 0 & 0 & 0 & 0 & 0 & 0 & 0 & 0 & 0  & 0 & 0  & 0  & 0 & 0 & 0 \\
Github Personal Access Token    & 0 & 0 & 0 & 0 & 0 & 0 & \textbf{1} & 0 & 0  & 0 & 0  & 0  & 0 & 0 & 0 \\ \midrule
Total    & 1 & 2 & \textbf{8} & 0 & \textbf{5} & \textbf{5} & 3 & 4 & \textbf{17} & 1 & 29 & \textbf{50} & \textbf{2} & 1 & 1 \\ \bottomrule
\end{tabular}
}\vspace{-0.4cm}
\end{table}
}
\vspace{-0.2cm}
\section{Introduction}
Large Language Models~(LLMs) have revolutionized the field of natural language processing, enabling groundbreaking advancements in various domains. A notable sub-domain of LLMs is Code LLMs, which specialize in generating, understanding, and manipulating programming code. Code LLMs have found extensive applications in code completion, code generation, bug fixing, and code summarization, demonstrating their significant impact on software development practices and productivity~\cite{xu2022systematic,niu2023empirical,zhou2024out,tu2023isolating,vaithilingam2022expectation}.
Despite the remarkable capabilities of LLMs, they raise a significant privacy concern. Specifically, LLMs are trained on vast amounts of data sourced from a wide range of locations, including public repositories, forums, and websites. This comprehensive training process inadvertently leads to the memorization of sensitive and private information present in the training data~\cite{meli2019bad}. Consequently, Code LLMs become prone to leaking sensitive and critical information~(e.g., user credentials and secrets) during the code generation process, posing severe privacy risks and breaches.

Existing studies on memorization in LLMs have primarily employed prompt engineering techniques to elicit the memorized information from the models~\cite{niu2023codexleaks, huang2023not}. While these approaches have provided valuable insights, they suffer from two main limitations. First, forcing or eliciting the LLM to output secret information might lead to widespread hallucination~\cite{su2024mitigatingentitylevelhallucinationlarge}, where the LLM generates false secrets, resulting in a high number of false positives. This can significantly hinder the accurate assessment of the true extent of memorization and privacy leakage risks. Second, the prompt engineering based approaches generat output randomly, which can be inefficient in extracting the target sensitive information memorized by LLMs. This randomness may lead to repeated extraction of similar secrets across multiple queries while missing other unique secrets, potentially underestimating the true extent and variety of sensitive information memorized by LLMs and hindering an effective assessment of the associated privacy risks. To fully grasp the intricacies of memorization in LLMs and overcome these limitations, it is imperative to delve deeper into the internal workings of these models and examine the issue at a more granular level.

\noindent\textbf{Our Work.} In this paper, we present a novel approach to characterize real and fake secrets generated by Code LLMs based on token probabilities. Through extensive observation and analysis, we have identified several key characteristics that distinguish genuine secrets from hallucinated ones, such as the stabilization of real secret tokens at high probabilities (C1), higher overall probabilities of real secret tokens compared to fake ones (C2), more pronounced probability advantages of real secret tokens (C3), and the ability to identify certain fake secrets early in the decoding process using secret strength metrics like Shannon entropy (C4). While we illustrate these characteristics using Google API Keys, it is important to note that they are generalizable to various types of secrets.

To overcome the aforementioned limitations from existing works in the literature, we propose a method named \tool that leverages token-level features derived from the identified characteristics to guide the token decoding process. \tool consists of two stages: (1) Offline Token Scoring Model Construction, which utilizes a proxy Code LLM to generate training data and train a scoring model to predict the likelihood score that a token belongs to a real secret, and (2) Online Scoring Model Guided Decoding, which leverages the scoring model to predict a score for the tokens at each decoding step, combining it with the original LLM-predicted probability to reassign token likelihoods and guide the selection of tokens. 

 Through extensive experiments on five state-of-the-art Code LLMs (StableCode, CodeGen2.5, DeepSeek-Coder, CodeLlama, and StarCoder2) and a diverse dataset of 1200 code files containing various types of secrets across multiple programming languages, we demonstrate the superior performance of \tool compared to existing baselines. \tool achieves an average Plausible Rate (PR) of 44.74\% across the five Code LLMs, significantly outperforming HCR (10.98\%) and BS-5 (23.60\%). Moreover, \tool successfully extracts a total of 1076 real secrets from the victim models, surpassing the numbers obtained by HCR (230 real secrets) and BS-5 (845 real secrets). These results highlight the effectiveness of our token-level approach in guiding the decoding process to generate more diverse and accurate secrets, enabling a more comprehensive assessment of the privacy leakage risks associated with Code LLMs.

 \noindent\textbf{Contributions.} We summarize the contributions as follows:
 \begin{itemize}[leftmargin=*]
 \item \textbf{Identification of key characteristics distinguishing real and fake secrets.} We identify four generalizable key characteristics that differentiate genuine secrets from hallucinated ones generated by Code LLMs, based on token probabilities and secret strength metrics. 

 \item \textbf{Development of a token-level approach for secret extraction.} We propose \tool, a two-stage method that constructs an offline token scoring model using a proxy Code LLM and employs it to guide the decoding process by reassigning token likelihoods based on token-level features derived from the identified characteristics.

 \item \textbf{Extensive evaluation on state-of-the-art Code LLMs.} We evaluate \tool on four state-of-the-art Code LLMs using a diverse dataset, demonstrating its superior performance in achieving a higher plausible rate and extracting more real secrets compared to existing baselines.
 \end{itemize}

 \noindent\textbf{Ethical Considerations.} 
In conducting this study, we prioritized ethical considerations to ensure responsible research practices. We strictly limited our use of the derived secrets to validating their authenticity without exploiting or misusing them in any way. Furthermore, we have taken great care to mask all the information presented in this paper, including any personal or sensitive data, to prevent potential leakage and real damage. By adhering to these ethical principles, we aim to contribute to the understanding of memorization in Code LLMs without causing any unintended consequences.

\section{Background and Related Work}\label{sec:background}

\subsection{Code LLMs}\vspace{-0.1cm}

Code LLMs, based on large language model technology and trained on vast code datasets, are designed to understand and generate programming code\cite{hou2023large}. They show significant potential in tasks such as code generation, autocompletion, and error detection\cite{xu2022systematic,niu2023empirical,zhou2024out,tu2023isolating,vaithilingam2022expectation}. Researchers are dedicated to enhancing their capabilities~\cite{liu2023codegen4libs,gao2024learning,liu2023empirical,zhou2024out,shi2024greening,zhou2024calibration}, leading to the emergence of various Code LLMs, including Codex\cite{chen2021evaluating}, CodeParrot\cite{tunstall2022natural}, CodeBERT\cite{feng2020codebert}, CodeGPT\cite{lu2102codexglue}, and Codellama\cite{roziere2023code}. These models have become indispensable tools for software developers worldwide, with applications such as Google's CodeSearchNet\cite{husain2019codesearchnet} using CodeBERT\cite{feng2020codebert} for code search, Visual Studio Code integrating GitHub Copilot\cite{copilot} and CodeGeeX\cite{zheng2023codegeex} as extensions, and Kaggle's AI Code Competition using Codex to help beginner programmers learn to code.

\vspace{-0.2cm}
\subsection{Memorization in LLMs}
\vspace{-0.1cm}
LLMs can remember a large amount of training data during the training process. This memorization mechanism enables the model to reproduce information from the training data when generating text. For a large model \( f \) trained on a dataset \( D \), given an input sequence \( x = (x_1, x_2, \ldots, x_n) \), the model generates an output sequence \( y = (y_1, y_2, \ldots, y_m) \). The parameters of the model \( f \) are denoted as \( \theta \).

If there exists a prompt \( p \) such that: $f(p, \theta) = s, s \in D$, then the model \( f \) is said to have memorized the string \( s \). Since \( s \) strictly exists in the training set \( D \), this phenomenon is summarized as Exact Matching in the study\cite{neel2023privacy}.
Additionally, Lee et al.\cite{lee2021deduplicating} defined a more lenient phenomenon called Approximate Matching: for \( f(p, \theta) = s' \), if there exists a corresponding string \( s \) in the training set \( D \) such that \( s \) and \( s' \) are within a specified edit distance, then the model \( f \) is also said to have memorized the string \( s \).

Many existing works have explored memorization in LLMs. Feldman\cite{feldman2020does} proposed that memorization of labels is necessary for near-optimal generalization error in data following natural distributions. Studies\cite{carlini2019secret,carlini2021extracting,parikh2022canary} reveal that attackers can perform training data extraction attacks by querying LLMs. Carlini et al.\cite{carlini2022quantifying} investigated factors influencing LLM memorization, finding that increases in model capacity, data repetition, and prompt length significantly enhance memorization.
Memorization also exists in Code LLMs, with studies\cite{ciniselli2022extent,rabin2023memorization} confirming that code models can memorize their training data. Yang et al.\cite{yang2023code} conducted a systematic study on memorization in code models, discussing factors such as data repetition and model size. For Multimodal Large Language Models, Chen et al.\cite{chen2024we} developed metrics to measure data leakage during multimodal training.

The memorization in Code LLMs leads to privacy issues such as the reproduction of sensitive information and the leakage of proprietary code. Al-Kaswan et al.\cite{al2023targeted} proposed a targeted attack to identify whether a given sample in the training data can be extracted from the model. Niu et al.\cite{niu2023codexleaks} explored the impact of temperature values on the generation of private information by GitHub Copilot, finding that about 8\% of the prompts resulted in privacy leaks. Huang et al.\cite{huang2023not} proposed HCR, in which Copilot generated 2702 hard-coded credentials and CodeWhisperer produced 129 keys during code completion tasks. Yao et al.\cite{wan2024does} introduced CodeMI, a membership inference method that determines whether a given code sample is present in the training set of a target black-box model by analyzing the output features of shadow models.

\vspace{-0.1cm}
\section{Token-Level Secret Characterization}\label{sec:study}

We initially conduct a study to investigate the characteristics of real secrets and fake secrets generated by Code LLMs at token level, aiming to answer the research question:

 \noindent\textbf{RQ1:} What are the token-level characteristics of secret memorization in Code LLMs?
\vspace{-0.2cm}
\subsection{Motivation}
While existing researches have shown that Code LLMs can memorize secrets or privacy from training datasets \cite{biderman2024emergent,carlini2022membership,ishihara2023training}, the characteristics of these memorized secrets remain under-explored. Additionally, given that Code LLMs may generate fake secrets due to hallucination issues \cite{huang2023not}, it is necessary to explore the differences between the characteristics of real and fake secrets~\cite{fadeeva2024fact}. 

To this end, we conduct an in-depth analysis of secret memorization in Code LLMs at the token level. \revise{We aim to extract the 6 most common types of commercial secrets included in the prior study~\cite{huang2023not}, which together constitute 85.8\% of all secrets. For ease of management, these secrets follow structured formats that can be defined using regular expressions. Table~\ref{tab:secret-types} lists these 6 types of secrets along with their corresponding regular expressions.}

\SecretTypes
\vspace{-0.2cm}
\subsection{Study Setup}\label{sec:study-steup}
We construct code completion prompts and input them into Code LLMs to predict the missing secrets based on these prompts. During this process, we collect the predicted token probabilities at each decoding step and analyze their characteristics. 

\subsubsection{Source File Collection}\label{sec:file-collection}
For the 6 types of secrets, we use their regular expressions to search for code files containing them from open-source repositories through SourceGraph~\cite{sourcegraph}. We choose SourceGraph for two reasons: (i) It integrates a vast number of open-source code repositories;

(ii) It supports efficient code search using regular expressions, which perfectly meets our requirement for searching secrets.

In the search process, we only consider 5 languages, namely HTML, Java, JavaScript, PHP, and Python, which are commonly used in web applications that may require the 6 types of secrets for online API requests. After duplication, 31,978 code files are identified \revise{as of June 16, 2024.} To ensure data quality, we apply the filtering rules adopted in the StarCoder project \cite{li2023starcoder} . Files that meet the following criteria are excluded:

\begin{itemize}[leftmargin=*]

   \item Files with an average line length exceeding 100 characters or a maximum line length exceeding 1000 characters.
   
   \item Files with less than 25\% alphabetic characters.
   
   \item Files containing the string ``\texttt{\textless ?xml version=}'' within the first 100 characters.
   
   \item For HTML files, we filter out files with less than 20\% visible text or fewer than 100 characters of visible text.
\end{itemize}

These filtering operations remove approximately 3.85\% of the files. Subsequently, we perform an additional examination of the characters surrounding the matched content to filter out the mismatched code files. \revise{For example, the middle part (highlighted in blue) of ``...kBd\textcolor{blue}{AIza...}BVV...'' matches the regular expression for Google API Key, but the surrounding characters indicate that it is not a valid Google API Key.}  
Ultimately, 66.57\% of the original search results are retained, resulting a set of 25,353 code files. Table \ref{table:raw dataset} shows the language and secret distribution in the final results.

\RawDataset

We randomly select 200 code files for each type of secret, totaling 1,200 files, for subsequent method evaluation (Section~\ref{sec:eval-setup}). The proportion of each programming language within each type of secret remains consistent with its proportion in the raw data. As a result, 24,133 code files are retained for analysis in this study.

\subsubsection{Completion Prompt Construction}\label{sec:prompt-construction}

From the remaining code files, we randomly sample 400 files for each type of secret to construct completion prompts. Note that the numbers of files for STSK, SIWU, TCSI, and ACAK are fewer than 400, so we additionally collect files in other languages like Go and C using SourceGraph to reach 400. In total, 2,400 code files are used to construct completion prompts.

\begin{figure}
    \centering
    \includegraphics[width=0.85\linewidth]{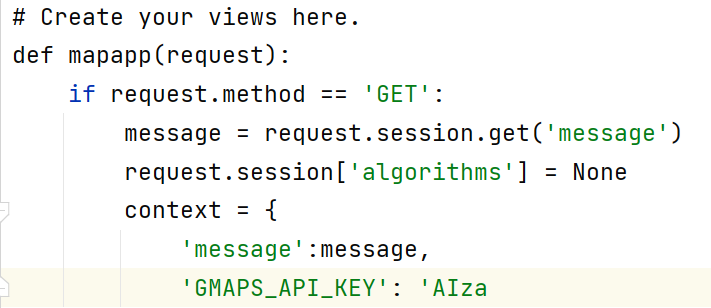}
    \caption{An Example of Completion Prompt}\vspace{-0.6cm}
    \label{fig:prompt_example}
\end{figure}

For each code file, we locate the secret's position and remove all content after the fixed secret prefix (\eg ``\texttt{AIza}'' for Google API Key). We then retain the line containing the secret and up to 7 preceding lines as the completion prompt, \revise{which provides sufficient context without overwhelming the model based on the preliminary observations and previous study \cite{beltagy2020longformer}.} For example, as illustrated in Fig.~\ref{fig:prompt_example}, the fixed prefix ``\texttt{AIza}'' of a found Google API Key is retained, while all subsequent content is removed. After processing all code files, we obtain 2,400 completion prompts.

\subsubsection{LLM-based Secret Generation}
We feed these 2,400 prompts into a \textbf{\textit{Proxy Code LLM}} to perform the code completion and predict the missing Google API Keys. In this study, we select StarCoder2-15B as the proxy LLM, which is a 15B-parameter model trained on multiple programming languages from The Stack v2 \cite{lozhkov2024starcoder}, supporting various tasks such as code completion and code generation. 
\revise{StarCoder2-15B is chosen as the proxy LLM because it shares the same common decoder-only architecture as the other Code LLMs in our study. This ensures consistency in studying secret token generation.}

StarCoder2-15B generates a token sequence step by step based on an input prompt. At each step, it generates a token probability distribution based on the context provided by the prompt and the previously predicted tokens. To avoid generating erroneous secrets that violate the format constraints in Table~\ref{tab:secret-types}, we set the probabilities of invalid tokens to 0. We then apply beam search with a width of 5 to select the next token, keeping the token sequence with the highest likelihood score as the final generated secret. \revise{The generation stops once the target secret's required length (character number) is reached.} During the generation process, we record the predicted token probability distribution of each step for further analysis.

\subsubsection{Generated Secret Verification}\label{sec:verification}
For each generated secret, we validate it through a combination of online API request and GitHub code search as follows:
\begin{itemize}[leftmargin=15pt]
    \item \textbf{API Request:} We develop verification scripts that send connection requests to the relevant services using the secret and observe the returned status codes. If the services are connected successfully, the secret is real and still active.
    \item \textbf{GitHub Search:} If the secret cannot be validated through an API request, we use GitHub Code Search\cite{github_code_search} to search for it. If it can be found through this search and is not merely a usage example (\eg ``\texttt{AKIDz8...EXAMPLE}''), it is also considered a real memorized secret.
\end{itemize}

The verified dataset consists of 622 secrets: 311 real and 311 fake. It includes 100 real and 100 fake secrets each for Google API Key, Stripe Test Secret Key, and Tencent Cloud Secret ID (totaling 600), plus 11 real and 11 fake secrets for Google OAuth Client ID (due to limited availability). Slack Incoming Webhook URLs and Alibaba Cloud Access Key IDs are excluded as the model does not generate real secrets for these types.

\vspace{-0.2cm}
\subsection{Characterization} 
Based on our observation and analysis of token probabilities, we characterize the real and fake secrets generated by the Code LLM as follows. Note that these characteristics are generalizable to different types of secrets, we simply use Google API Keys to illustrate (see our replication package~\cite{Replication} \revise{for STSK, TCSI, and GOCI}).

\begin{figure}
    \centering
    \includegraphics[width=0.95\linewidth]{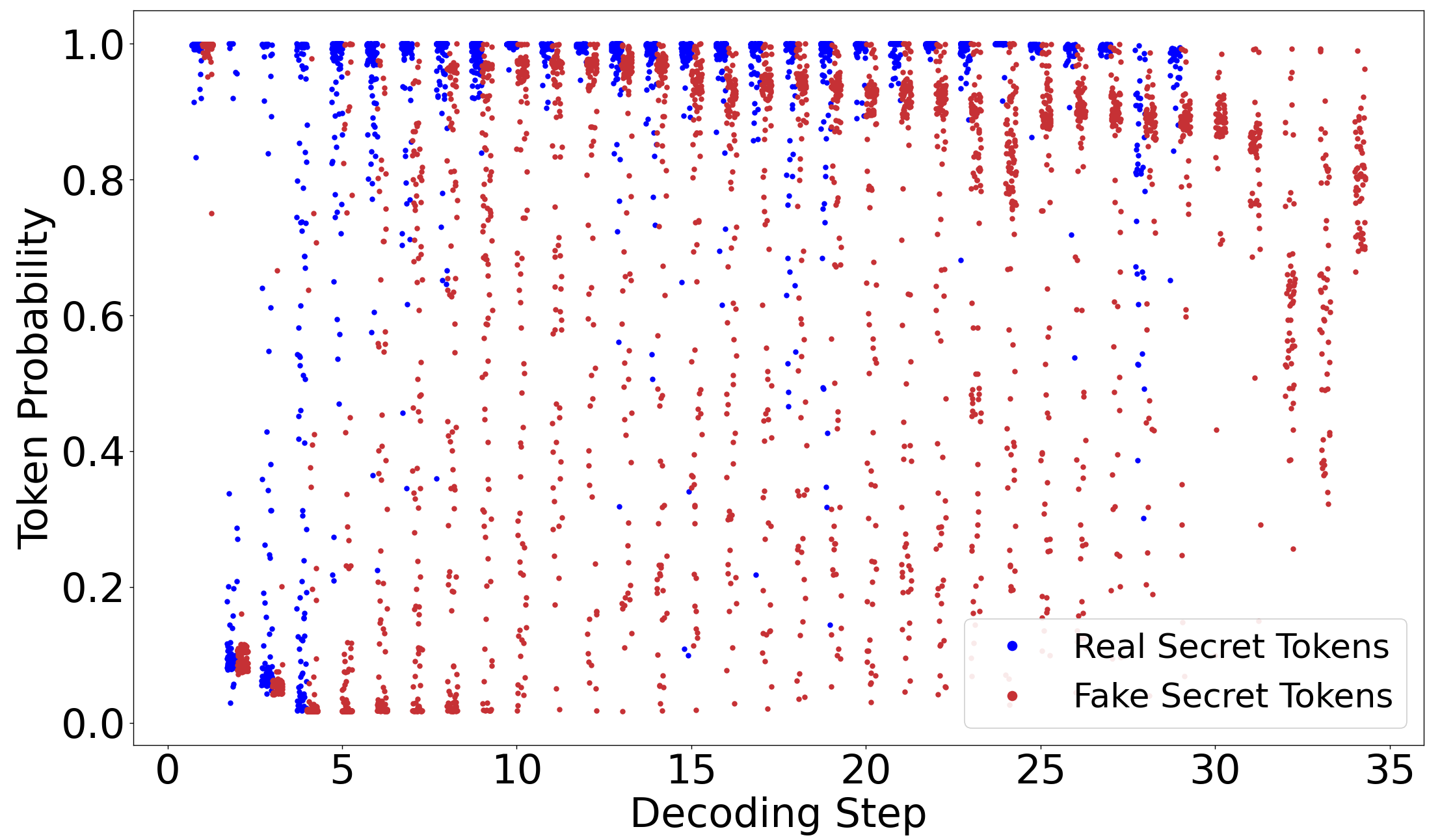}
    \vspace{-0.2cm}
    \caption{Token Probability Scatter Plot for Google API Keys}
    \vspace{-0.5cm}
    \label{fig:token_prob_scatter_plot}
\end{figure}

\begin{figure}
    \centering
    \includegraphics[width=0.95\linewidth]{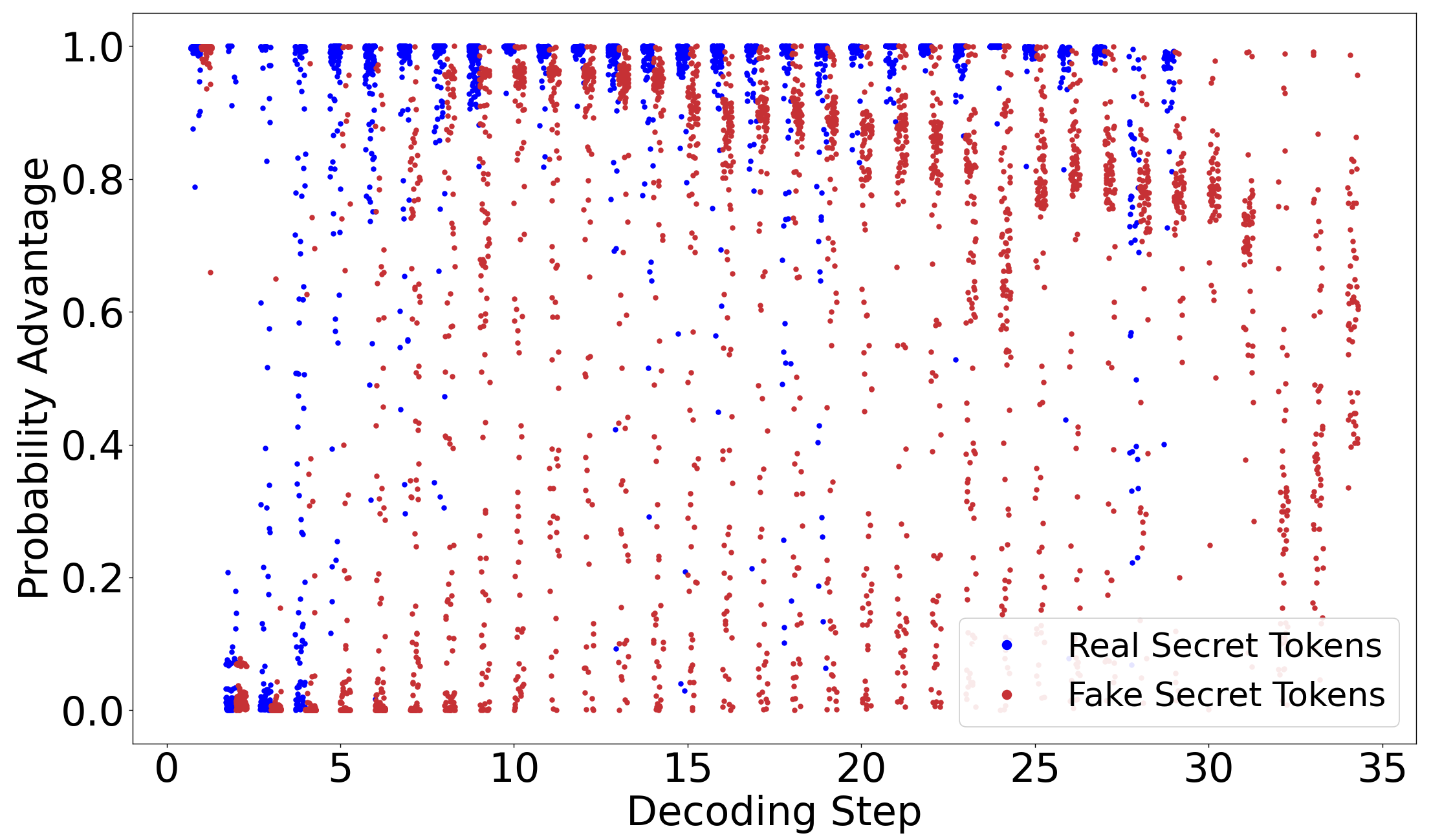}
    \vspace{-0.2cm}
    \caption{Token Probability Advantages Scatter Plot for Google API Keys}
    \vspace{-0.8cm}
    \label{fig:Prob_distance scatter_plot}
\end{figure}

\textbf{C1: Tokens in real secrets tend to stabilize at relatively high probabilities after initial few decoding steps.}
Fig. \ref{fig:token_prob_scatter_plot} shows the scatter plot of token probabilities for both real and fake secrets (\textcolor{blue}{blue}  and \textcolor{red}{red} points respectively) along the decoding steps, where each point ($i$, $prob$) represents a token at the $i$-th decoding step with a probability of $prob$. \revise{Typically, the probabilities of the initial few steps (e.g., the first five) for real secret tokens are relatively low. However, as the decoding progresses, the probability values rapidly increase and stabilize at a relatively high level.} For instance, at the 10th step, tokens of the 100 real secrets have probabilities over 0.9. \revise{This observation can be attributed to the model's initial uncertainty in earlier steps; however, as it decodes more tokens and aligns with memorized sequences, its confidence increases quickly. Besides, an interesting discovery is that real secret tokens typically consist of multiple characters, while fake ones often have one single character. Consequently, real secrets decode in fewer steps and terminate earlier.}

\textbf{C2: Tokens in real secrets generally have higher probabilities than those in fake ones.} 
According to Fig. \ref{fig:token_prob_scatter_plot}, the overall probabilities of real secret tokens are generally higher than those of fake secret tokens, with the latter often more scattered between 0 and 1. For example, at the 10-th step, the probability values of many fake secret tokens are evenly distributed between 0 and 0.9. \revise{This can be explained by the internal memory mechanism of LLMs. In addition, we conducted t-tests for this characteristic, and the results indicate a significant probability difference between real tokens and fake tokens for all secret types ($p \ll 0.01$).}

\textbf{C3: Probability advantage of tokens in real secrets is typically more pronounced than in fake ones.} The probability advantage of a token at a given decoding step is defined as the difference between its probability and the probability of the next token in the distribution, ranked by descending probabilities. Fig. \ref{fig:Prob_distance scatter_plot} shows a scatter plot of token probability advantages for both real and fake secrets along the decoding steps, where each point ($i$, $adv$) represents a token at the $i$-th decoding step with a probability advantage of $adv$. Similar to \textbf{C2}, the overall probability advantages of real secret tokens are generally higher than those of fake secret tokens. \revise{In addition, we conducted t-tests for this characteristic, and the results indicate a significant probability advantage difference between real tokens and fake tokens for all secret types ($p \ll 0.01$).}

\textbf{C4: Certain fake secrets can be identified early in decoding process using secret strength metrics like Shannon entropy.}
Code LLMs often generate weak secrets containing continuously repeated tokens due to the decoding objective of maximizing likelihood \cite{xu2022learning}. As the repetition of characters or tokens increases during the decoding process of such fake secrets, the Shannon entropy of the generated token sequence continuously decreases. This observation inspires us to \revise{continuously} inspect the Shannon entropy during the decoding process and avoid certain fake secrets at an early stage to increase the likelihood of generating real secrets. \revise{Although entropy is commonly used to assess the strength of secrets, we are the first to integrate it directly into the LLMs' decoding process, rather than as a post-processing step.}  For example, if the current decoding step token sequence is ``\texttt{SyA2-}'', and ``\texttt{2}'' and ``\texttt{x}'' are the two tokens with the highest probabilities predicted by the Code LLM, we can discard ``\texttt{2}'' and choose ``\texttt{x}'' as the next token since appending ``\texttt{2}'' would decrease the sequence's entropy, thus preventing Code LLMs from falling into token repetition and producing fake secrets like ``\texttt{AIzaSyA2-2-2-2...}''.


\begin{tcolorbox}[colback=gray!10, colframe=gray!40, title=Answer to RQ1, fonttitle=\bfseries\color{black}]
\vspace{-0.1cm}
This study reveals that real and fake secrets generated by Code LLMs exhibit different token-level characteristics (\textbf{C1}-\textbf{C4}). This insight motivates us to leverage token-level features to improve the extraction of secrets memorized in Code LLMs.
\vspace{-0.1cm}
\end{tcolorbox}

\section{Methodology}\label{sec:methodology}

\begin{figure*}
    \centering
    \includegraphics[width=0.88\linewidth]{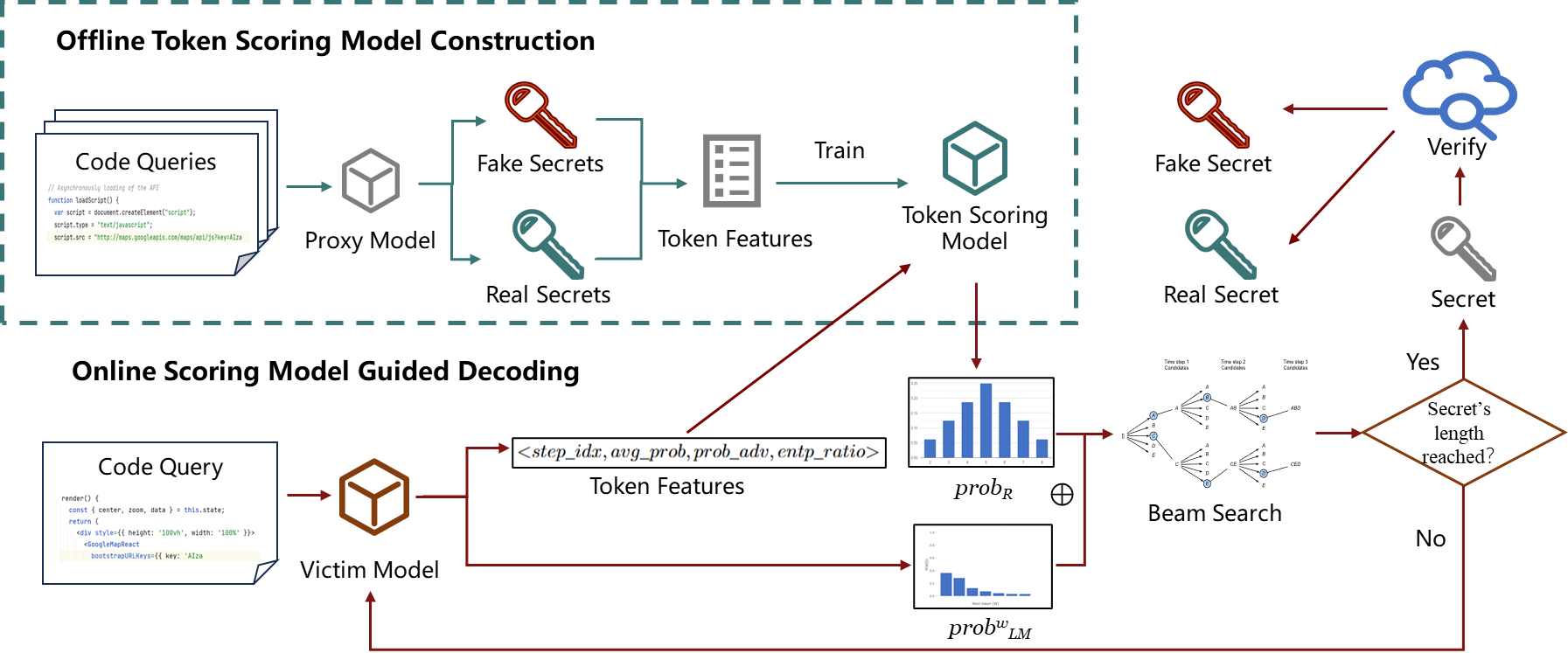}
    \caption{Overall Workflow of \tool}\vspace{-0.5cm}
    \label{fig:pipeline}
\end{figure*}

Based on our characterization, we propose \tool~(see Fig. \ref{fig:pipeline} for an overview), a method to extract secrets from Code LLMs by guiding the token decoding process with token-level features derived from \textbf{C1}, \textbf{C2}, \textbf{C3}, and \textbf{C4}. These features capture the characteristics of tokens that help determine whether they belong to a real secret.
\tool consists of two stages: (1) \textbf{Offline Token Scoring Model Construction}, which utilizes a proxy Code LLM to generate training data and train a scoring model to predict the likelihood score that a token belongs to a real secret; and (2) \textbf{Online Scoring Model Guided Decoding}, which leverages the scoring model to predict a score for the tokens at each decoding step. The score is then combined with the original LLM-predicted probability to reassign the token likelihoods, guiding the selection of tokens.

\vspace{-0.4cm}
\subsection{Token-Level Features}\label{sec:features}
\vspace{-0.1cm}
\begin{figure}
    \centering
    \includegraphics[width=1\linewidth]{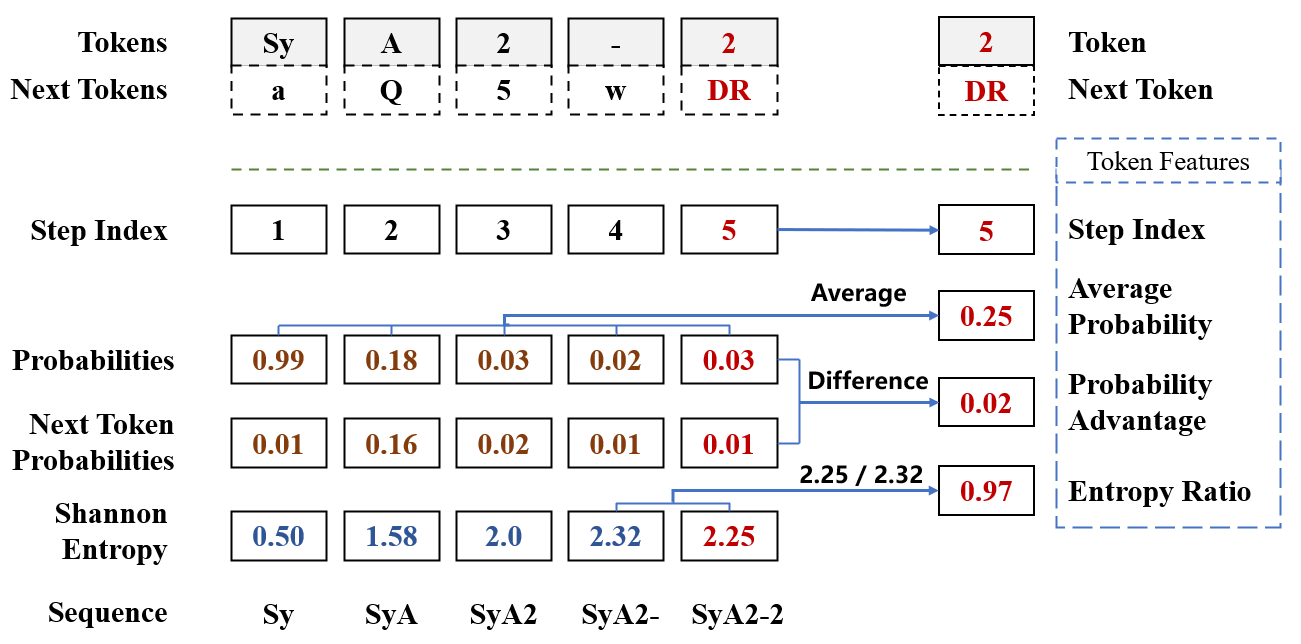}
    \caption{Token Features Extraction Process}\vspace{-0.7cm}
    \label{fig:token_features}
\end{figure}

We first define four features for each token during the decoding process of Code LLMs, as follows:

\textbf{Step Index ($step\_idx$)}: According to \textbf{C1}, real secret tokens exhibit distinct probability distributions at different decoding steps. To capture this, we use the step index of a token in the decoding process as a feature. For example, in Fig. \ref{fig:token_features}, the token ``\texttt{2}'' generated at step 5 has a Step Index of 5.

\textbf{Average Probability ($avg\_prob$)}: 
According to \textbf{C2}, real secret tokens generally have higher probabilities than fake ones. To capture this while addressing the instability of using a single token's probability, we use the average probability of tokens selected in previous decoding steps as a feature. \revise{Compared with token's probability sequence, the average probability could reduce computational complexity while ensuring interpretability (see our replication package \cite{Replication}).} In Fig. \ref{fig:token_features}, the probabilities of tokens up to ``\texttt{2}'' are 0.99, 0.18, 0.03, 0.02, and 0.03, with an average of 0.25, which is set as the Average Probability for token ``\texttt{2}''.

\textbf{Probability Advantage ($prob\_adv$)}: According to \textbf{C3}, a token's probability advantage often indicates whether it belongs to a real secret. We capture this as a feature by calculating the difference between the token's probability and that of the next token in the probability distribution. In Fig. \ref{fig:token_features}, at step 5, the probability of token ``\texttt{2}'' is 0.03, and the next token ``\texttt{DR}'' has a probability of 0.01. Thus, the Probability Advantage for token ``\texttt{2}'' is set to 0.02.

\textbf{Entropy Ratio ($entp\_ratio$)}: According to \textbf{C4}, the changing trend of Shannon entropy in the predicted sequence during decoding can help filter out fake secrets early. We capture this by using the ratio of the current step's entropy to the previous step's entropy as a feature. In Fig. \ref{fig:token_features}, for the token ``\texttt{2}'' generated at the current step, the corresponding string is ``\texttt{SAy2-2}'' with a Shannon entropy of 2.25. The previous step's string was ``\texttt{SAy2-}'' with a Shannon entropy of 2.32. Thus, the Entropy Ratio for token ``\texttt{2}'' is set to 0.97 ($2.25/2.32$).

\textbf{\emph{Feature Vectorization.}} We form a feature vector for a token based on its four features as follows:
\vspace{-0.3cm}
$$\mathbf{feat}=\textless step\_idx, avg\_prob, prob\_adv, entp\_ratio\textgreater $$
\vspace{-0.6cm}

For example, the feature vector for the token ``\texttt{2}'' in Fig. \ref{fig:token_features} is:
\vspace{-0.3cm}
$$ \mathbf{feat}=\textless 5, 0.25, 0.02, 0.97\textgreater $$

\vspace{-0.2cm}
\subsection{Token Scoring Model Construction}
\vspace{-0.1cm}
To combine the four independent features and assess the probabilities of tokens belonging to real secrets, we train a token scoring model.

\textbf{Training Data Construction.}
We follow a process similar to our characterization study setup (Section~\ref{sec:study-steup}). We create completion prompts for each secret type based on searched code files and feed them into a \textbf{\textit{Proxy Code LLM}} to generate candidate secrets. The generated candidates are validated and split into a real set $\mathcal{R}$ and a fake set $\mathcal{F}$. During the proxy Code LLM's token decoding process, we extract the feature vector $\mathbf{feat}$ for each token, resulting in two vector sets: $\mathbf{Feat}_\mathcal{R}$ and $\mathbf{Feat}_\mathcal{F}$, corresponding to $\mathcal{R}$ and $\mathcal{F}$, respectively.

\textbf{Scoring Model Training.}
To effectively distinguish between tokens of real and fake secrets, we train a Linear Discriminant Analysis (LDA) \cite{hart2000pattern} model to find a combination of the four features that maximizes the differentiation between $\mathbf{Feat}_\mathcal{R}$ and $\mathbf{Feat}_\mathcal{F}$. 
We choose this linear model over complex models like Multi-Layer Perceptron (MLP) \cite{riedmiller2014multi} 
because our feature vector for each token contains only four features, and using non-linear models can lead to overfitting issues with low-dimensional feature vectors.

\revise{We prioritize the resource-efficient LDA rather than other complex models due to the immediacy required in decoding.}

After training on $\mathbf{Feat}_\mathcal{R}$ and $\mathbf{Feat}_\mathcal{F}$, the resulting combination can be used by the scoring model to predict the probability that a feature vector $\mathbf{feat}$ belongs to real secrets:
\vspace{-0.2cm}
\begin{equation}\label{eq:scoring}
    prob_\mathcal{R} = \mathbf{SM}(\mathbf{feat})
\end{equation}

\vspace{-0.4cm}
\subsection{Scoring Model Guided Decoding}
\vspace{-0.1cm}
Using the scoring model, we guide decoding process of the \textit{Victim Code LLM}, as detailed in Algorithm~\ref{alg:decoding}, to enhance the likelihood of generating real secrets. The algorithm follows the traditional beam search decoding process, with additional enhancement steps (highlighted in \textcolor{red!60!black}{red}) including \textit{Masking Invalid Tokens} (line 6), \textit{Extracting Token Features} (line 9), \textit{Calling Scoring Model} (line 10), and \textit{Combining Probabilities} (line 11). The last three additional steps constitute the process referred to as \textit{Scoring Model Guided Probability Re-weighting}. We first briefly introduce the overall process based on beam search and then describe the goals and the details of the enhancement steps.

\begin{algorithm}
\caption{Scoring Model Guided Decoding}
\label{alg:decoding}

\small
\DontPrintSemicolon
    \KwData{Prompt $pmpt$, Beam Size $B$, Length Limit $L$}
    \KwResult{Secret $sec$}
    $Beam \gets [(pmpt, 0)]$ \tcp*{initialize hypothesis pool}
    $step \gets 0$ \\
    \While{character \# of sequences in Beam $< L$}{
        $step \gets step + 1$ \\
        $Cands \gets []$ \tcp*{initialize candidate list}
        \ForEach{$(seq, score) \in Beam$}{
            $\mathbf{p} \gets \mathbf{LM}(seq)$ \tcp*{get probabilities for next token}
            \textcolor{red!60!black}{$\mathbf{p} \gets \textsc{MaskInvalidToks}(\mathbf{p})$} \tcp*{mask invalid tokens}
            $Toks \gets \textsc{TopK}(\mathbf{p}, B)$ \tcp*{get top-B tokens by probability}
            \ForEach{$(tok, prob_{LM}) \in Toks$}{
                \textcolor{red!60!black}{$\mathbf{feat} \gets \textsc{ExtractFeats}(tok, step, seq, \mathbf{p})$} \tcp*{extract features}
                \textcolor{red!60!black}{$prob_{\mathcal{R}} \gets \textbf{SM}(\mathbf{feat})$} \tcp*{call scoring model}
                \textcolor{red!60!black}{$prob \gets prob_{LM}^{w} \times prob_{\mathcal{R}}$} \tcp*{combine probabilities}
                $seq \gets seq \oplus tok$ \\
                $score \gets score + \log(prob)$ \\
                $Cands.\text{append}((seq, score))$
            }
        }
        $Cands \gets \text{rank}(Cands)$ \tcp*{rank hypotheses by score}
        \If{$step \leq K$} {
            $Beam \gets Cands$ \\
            \textbf{continue}
        }
        $Beam \gets Cands[:B]$ \tcp*{select top-B hypotheses}
    }
    $sec \gets \textsc{Argmax}(Beam)$ \tcp*{select the best hypothesis}
    \Return $sec$ 

\end{algorithm}

\textbf{Overall Beam Search Process.} The overall decoding process follows the key steps of beam search:
\begin{itemize}[leftmargin=*]
    \item \textbf{\textit{Hypothesis Pool Initialization} (line 1).} First, the process initializes a pool $Beam$ to maintain multiple hypotheses, where each hypothesis is a pair of $(seq, score)$, consisting of a token sequence $seq$ and a likelihood score $score$. At the beginning, there is only one hypothesis $(pmpt, 0)$ in $Beam$, where $pmpt$ is the input prompt.
    
    \item \textbf{\textit{Hypothesis Expansion} (lines 5-16).} For each hypothesis $(seq, score)$ in $Beam$, it is expanded to $B$ new hypotheses which are added to a candidate list $Cands$. Specifically, the process first selects the top $B$ tokens ($Toks$) from the LLM-predicted token probability distribution (lines 7-9) and then appends each selected token $tok$ to $seq$ (line 14), updating the likelihood score $score$ with $tok$'s log probability $\log(prob_{LM})$ (line 15).
    
    \item \textbf{\textit{Hypothesis Ranking and Pruning} (lines 17-21).} After expanding all hypotheses in $Beam$, there are $B$ times new candidate hypotheses in the candidate list $Cands$. The hypotheses are ranked based on their likelihood scores (line 17), and the top $B$ hypotheses are selected to update the hypothesis pool $Beam$ (line 21).
    
    \item \textbf{\textit{Final Secret Selection} (line 22).} These steps are iteratively performed until the length limit $L$ of the target secret type is reached. After that, the hypothesis with the highest likelihood score in $Beam$ is selected, and the token sequence is returned as the final secret.
\end{itemize}

\textbf{Our Enhancements.} We outline the goals and details of the enhancement steps as follows.
\begin{itemize}[leftmargin=*]
    \item \textbf{\textit{Masking Invalid Tokens} (line 8).} 
    In traditional beam search, the LLM predicts the probability distribution $\mathbf{p}$ based on the existing token sequence $seq$ (line 7), and the next step is to select the top $B$ tokens from this distribution (line 9). However, in our secret extraction scenario, some tokens with unsupported characters are invalid~(e.g., ``\texttt{*}''), as secrets must adhere to specific formats. To address this \revise{and avoid invalid candidates for the scoring model}, we mask invalid tokens in $\mathbf{p}$ by setting their probabilities to 0 (\ie the \textsc{MaskInvalidToks} procedure), ensuring they are not selected in the subsequent step. The specific constraint for determining invalid tokens for each secret type is listed as a regular expression in Table~\ref{tab:secret-types}.

    \item \textbf{\textit{Scoring Model Guided Probability Re-weighting} (line 11-13).} 
    In traditional beam search, each token $tok$ with its probability $prob_{LM}$ among the top $B$ tokens is used to expand a hypothesis (line 14-16). Our scoring model-guided decoding re-weights $prob_{LM}$ before constructing the new hypothesis using the following three steps:

    \begin{itemize}
        \item \textbf{\textit{Extracting Token Features} (line 11)}. We extract the feature vector $\mathbf{feat}$ for the token $tok$ using the \textsc{ExtractFeats} procedure, as outlined in Section~\ref{sec:features}.
        
        \item \textbf{\textit{Calling Scoring Model} (line 12).} Based on the extracted $\mathbf{feat}$, we call the trained scoring model to predict $prob_{\mathcal{R}}$ for $tok$ using Equation~\ref{eq:scoring}, indicating the probability that the given $tok$ belongs to a real secret \revise{under the condition that it is selected by LLM}.
        
        \item \textbf{\textit{Combining Probabilities} (line 13).} We combine the LLM-predicted probability $prob_{LM}$ and the scoring-model-predicted probability $prob_{\mathcal{R}}$ by multiplying them. The resulting probability ($prob_{LM}^{w} \times prob_{\mathcal{R}}$) represents a conditional likelihood, indicating the chance that $tok$ will complete a real secret when appended to the preceding tokens in $seq$. Note that $w$, within the interval [0,1], is a hyper-parameter used to control the weight of $prob_{LM}$.
    \end{itemize}
\end{itemize}

\textbf{Additional Optimization.} 
Based on \textbf{C1} from our characterization study, we optimize the beam search process (lines 18-20) by retaining all expanded candidate hypotheses ($Cands$) during the first $K$ steps instead of selecting the top $B$ hypotheses. This prevents missing real secrets due to the initially low probabilities of early tokens~\cite{9390407}. We set $K$ to 4 based on our observations in \revise{small-scale experiments and computational cost considerations.}

\vspace{-0.3cm} 
\section{Experimental Setup}\label{sec:eval-setup}
\vspace{-0.1cm}
We evaluate the effectiveness of our approach \tool through answering the following research questions.
\begin{itemize}[leftmargin=*]
    \item \textbf{RQ2.a:} How effective is \tool in extracting plausible secrets from Code LLMs? 
    \item \textbf{RQ2.b:} How effective is \tool in extracting real secrets from Code LLMs? 
    \item \revise{\textbf{RQ3.a:}} What are the effects of components of \tool on secret extraction?
    \item \revise{\textbf{RQ3.b:}} How effective is our token scoring model in identifying secret tokens?
    \revise{\item \textbf{RQ4:} How generalizable is \tool with respect to different Code LLMs and types of secrets?}
\end{itemize}

\vspace{-0.3cm}
\subsection{Evaluation Dataset}
As mentioned in Section~\ref{sec:file-collection}, we reserve 1,200 code files for method evaluation. Table \ref{table:dataset} details the number of each secret type and the programming language distribution in the dataset. We construct completion prompts based on these 1,200 code files using the same approach described in Section~\ref{sec:prompt-construction}.

\vspace{-0.2cm}
\Dataset

\vspace{-0.2cm}
\subsection{Victim Code LLMs}
We select the following open-source Code LLMs as victim models in addition to StarCoder: StableCode-3B~\cite{stable-code-3b}, CodeGen2.5-7B-multi~\cite{Nijkamp2023codegen2}, DeepSeek-Coder-6.7B-instruct~\cite{guo2024deepseek}, and CodeLlama-13B~\cite{roziere2023code}. These models vary in size, training data, and functionality, providing a diverse set of victim models for our experiments.

\vspace{-0.4cm}
\subsection{Baselines}
\vspace{-0.1cm}
We compare \tool with two baselines: (1) \textbf{HCR}~\cite{huang2023not}, a recent approach that reveals memorized secrets in neural code completion tools by constructing code infilling prompts, where the first line containing a secret is masked with a special token like \texttt{[MASK]} and other secrets are removed to eliminate the influence of context, and (2) \textbf{BS-5}, which directly applies beam search with size 5 using the same prompts as in \tool.

\vspace{-0.2cm}
\subsection{Metrics}
\vspace{-0.1cm}
\label{sec: metrics}
We evaluate the effectiveness of \tool and the baselines using two metrics:

 \begin{itemize}[leftmargin=*]
\item\textbf{Plausible Secrets}: Secrets that pass four filters—regex, entropy, pattern, and common words~\cite{huang2023not}—are considered plausible secrets. We use Plausible Rate (PR) to measure the ratio of the number of plausible secrets (PS\#) to the total number of generated secrets.

\item\textbf{Real Secrets}: Secrets that pass the verification process described in Section~\ref{sec:verification} are considered real secrets.
 \end{itemize}

\vspace{-0.2cm}
\subsection{\revise{Implementation}}\label{sub:implementation}
In the offline construction of the scoring model, we use StarCoder2-15B as the proxy. Since our proxy model does not generate real secrets for Slack Incoming Webhook URLs and Alibaba Cloud Access Key IDs, we use the token features of Google OAuth Client IDs and Stripe Test Secret Keys of corresponding lengths as the training dataset for training the scoring models of these two secrets.

\begin{figure}
    \centering
    \includegraphics[width=1\linewidth]{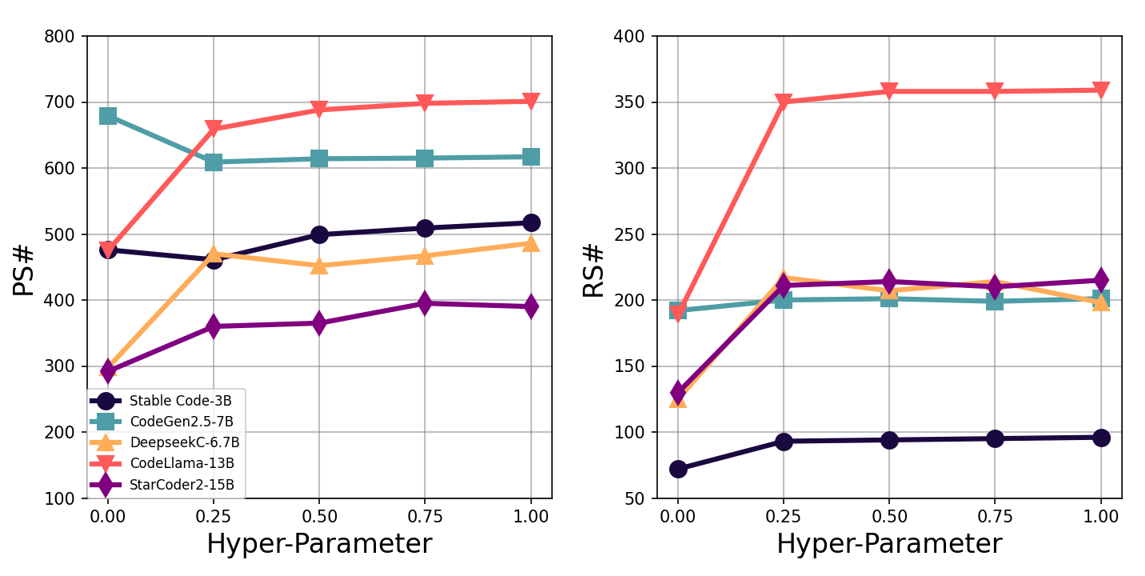}
    \vspace{-0.6cm}
    \caption{Trend of Plausible and Real Secret Count with Changes in Hyper-Parameter}
    \vspace{-0.5cm}
    \label{fig: hyper-parameters}
\end{figure}

In the online decoding, the beam search size $B$ is set to 5, \revise{balancing token search breadth and text quality~\cite{ott2018analyzing}.} The hyper-parameter \textit{w} for the LLM-predicted token probability is set to 0.75 by evaluating the results within interval [0,1]. 
Fig. \ref{fig: hyper-parameters} shows that it yields the highest number of real secrets and the best overall performance for generating both plausible secrets (\textbf{PS\#}) and real secrets (\textbf{RS\#}). 

\revise{Note that Token Scoring Model Construction and Scoring Model Guided Decoding involve no randomness, ensuring reproducibility; thus, we do not conduct repeated experiments.}

\textbf{Experimental Platform.} All the experiments are conducted on a server equipped with eight NVIDIA Tesla V100 SXM2 32GB NVLink GPUs. We verify the secrets generated by the models on a PC with an 11th Gen Intel(R) Core(TM) i7-11800H @ 2.30GHz CPU, 16GB DDR4 RAM, running Windows 11 as the operating system.
\section{Results and Analysis}
\subsection{RQ2.a: Effectiveness in Plausible Secret Extraction}
\begin{figure}[h]
    \centering
    \begin{subfigure}{0.48\textwidth}
        \centering
        \includegraphics[width=\textwidth]{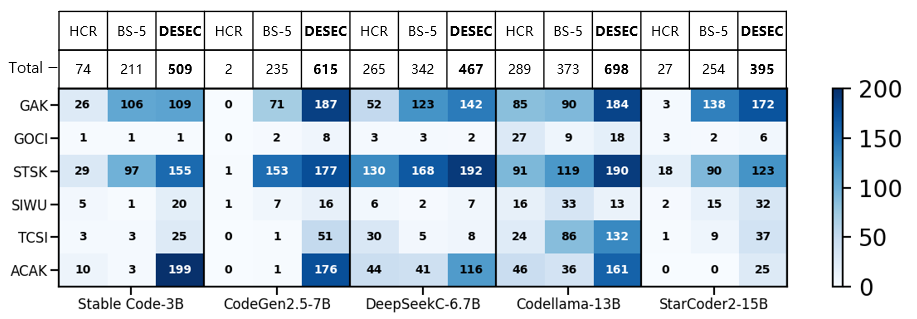}\vspace{-0.2cm}
        \caption{\textbf{PS\#} for extracting secrets}
        \label{fig: PS heat}
    \end{subfigure}
    \hfill
    \begin{subfigure}{0.48\textwidth}
        \centering
        \includegraphics[width=\textwidth]{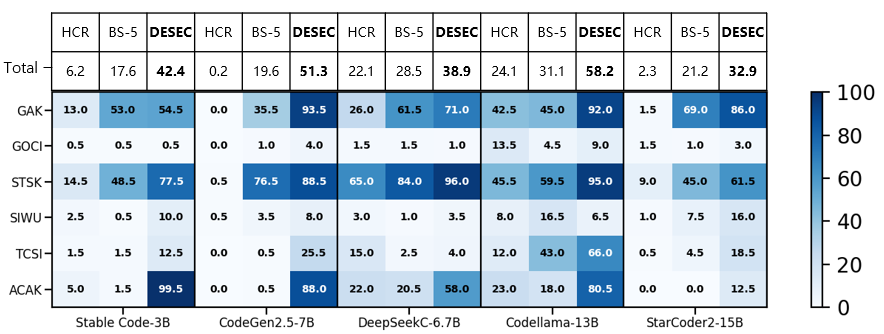}\vspace{-0.2cm}
        \caption{\textbf{PR\%} for extracting secrets}
        \label{fig: PR heat}
    \end{subfigure}
    \begin{subfigure}{0.48\textwidth}
        \centering
        \includegraphics[width=\textwidth]{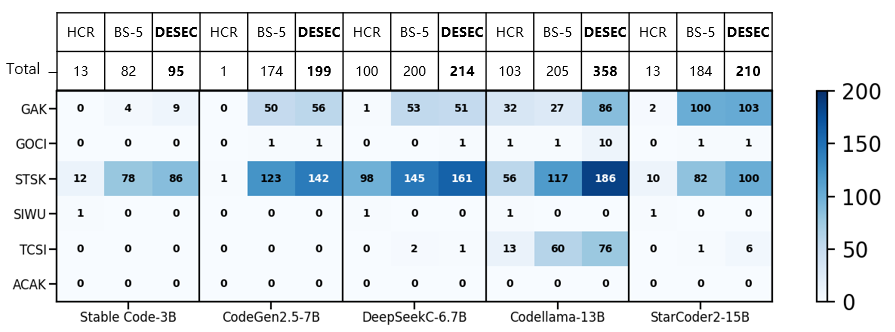}\vspace{-0.2cm}
        \caption{\textbf{RS\#} for extracting secrets}
        \label{fig: RS heat}
    \end{subfigure}\vspace{-0.2cm}
    \caption{\revise{Detailed experimental results for extracting secrets}}   \vspace{-0.3cm}
    \label{fig: detailed results}
\end{figure}

Fig. \ref{fig: PS heat} and Fig. \ref{fig: PR heat} present the detailed results of the effectiveness of \tool and the baselines in extracting plausible secrets, in terms of the number of plausible secrets (\textbf{PS\#}) and the plausible rate (\textbf{PR\%}).

\vspace{-0.5cm}

\textbf{Overall Results.} \tool effectively extracts 395-698 plausible secrets from the five victim Code LLMs, with plausible rates of 32.9-58.2\%.  However, the effectiveness of \tool and the two baselines varies significantly across different secret types. The discrepancy is due to two main factors: (i) the distribution of different types of secrets and (ii) their complexity. For example, \tool shows low effectiveness in extracting plausible GOCI secrets (\ie 0.5\% \textbf{PR\%}) while being relatively more effective for ACAK secrets (\ie 99.5\% \textbf{PR\%}). GOCI is the second most prevalent secret type in the collected source files (8,777 instances, see Table~\ref{table:dataset}), but its high complexity (12 digits + ``-'' + 32 characters) makes it challenging to extract. In contrast, ACAK has only 262 corresponding source files but is shorter (20 characters) and therefore easier to predict.

\textbf{Comparison to HCR.} 
\tool significantly outperforms HCR in extracting plausible secrets, with PR\% improvements ranging from 16.8\% to 51.2\% across the five victim Code LLMs. \tool outperforms HCR for most secret types, except for GOCI in DeepSeek-Coder-6.7B and CodeLlama-13B, where HCR extracts 1 and 9 more plausible secrets, respectively. The superiority of \tool can be attributed to: (i) \revise{The excessively long context of HCR distracts the model's attention\cite{beltagy2020longformer}, introducing noise and reducing focus on the target secret} 
, and (ii) beam search decoding in \tool being more effective than greedy search in HCR, evidenced by comparing HCR and BS-5 results.

\textbf{Comparison to BS-5.}  
Compared to BS-5, \tool demonstrates significant improvements in plausible secret extraction across all five models, with improvements ranging from 10.4\% to 31.7\%. The highest improvement of \tool is observed with CodeGen2.5-7B, where it extracts 380 more plausible secrets than HCR in total, indicating \tool's effectiveness in improving performance for CodeGen2.5-7B. \tool outperforms BS-5 for almost all secret types across the five models. Given that \tool is built on BS-5 (beam search with a size of 5), the improvements can be attributed to the enhancement of \textit{masking invalid tokens} introduced in the decoding process, which significantly reduces the likelihood of generating secrets that violate secret formats.

\begin{tcolorbox}[colback=gray!10, colframe=gray!40, title=Answer to RQ2.a, fonttitle=\bfseries\color{black}]
\vspace{-0.1cm}
\tool outperforms the baselines in prompting five victim Code LLMs to generate plausible secrets, with plausible rates ranging from 32.9\% to 58.2\%, due to the enhancement of \textit{masking invalid tokens} in the decoding process.
\vspace{-0.1cm}
\end{tcolorbox}

\vspace{-0.4cm}
\subsection{RQ2.b: Effectiveness in Real Secret Extraction}
Fig. \ref{fig: RS heat} presents the detailed results of the effectiveness of \tool and the baselines in extracting real secrets, in terms of the number of real secrets (\textbf{RS\#}).

\tool consistently demonstrates high effectiveness in extracting real secrets across all five victim Code LLMs, extracting between 95 and 358 real secrets, \revise{and reaches a total of 1076 (95+199+214+358+210), surpassing the numbers obtained by HCR (230 real secrets) and BS-5(845 real secrets).} 

The higher \textbf{RS\#} for CodeLlama-13B may indicate that it memorizes more real secrets, posing more severe privacy leakage risks.

Among all secret types, \tool and the baselines extract the highest number of STSK, which is the third most prevalent in the collected source files (569 instances, see Table~\ref{table:dataset}), likely due to its simple pattern and shorter length for testing purposes. However, \tool generates relatively few real GOCI, SIWU, and ACAK secrets across all Code LLMs.

\textbf{Comparison to HCR.} 
\tool significantly outperforms HCR in extracting real secrets (\textbf{RS\#}) across all five models, extracting 82 to 255 more secrets (as seen in the "Total" rows). HCR demonstrates inconsistent effectiveness, with very low \textbf{RS\#} for StableCode-3B, CodeGen2.5-7B, and StarCoder2-15B, indicating that its prompting and greedy decoding strategies may not be as generalizable as \tool across diverse Code LLMs.

\textbf{Comparison to BS-5.}
\tool outperforms BS-5 for most secret types, with the most significant improvement on CodeLlama-13B, resulting in a 74.6\% increase (153 more) in real secrets. Since \tool is built on BS-5, the improvements can be attributed to the \textit{scoring model guided probability re-weighting} enhancement, which leverages the four token-level features derived from the identified characteristics \textbf{C1}-\textbf{C4}.

\begin{tcolorbox}[colback=gray!10, colframe=gray!40, title=Answer to RQ2.b, fonttitle=\bfseries\color{black}]
\vspace{-0.1cm}
\tool consistently outperforms the baselines in prompting the five victim Code LLMs to generate real secrets, producing 95-358 real secrets across the models, due to its scoring model-guided decoding strategy that leverages the identified characteristics.
\vspace{-0.1cm}
\end{tcolorbox}
\vspace{-0.2cm}

\subsection{\revise{RQ3.a}: Ablation Study}\label{sec:ablation}
To investigate the effects of the two key enhancements in \tool, namely \textit{masking invalid tokens} and \textit{\revise{scoring model-guided probability re-weighting step}}, we conduct an ablation study. We respectively remove the two enhancements, resulting in two variants of \tool: \textbf{\textit{w/o masking}} and \textbf{\textit{w/o scoring}}.

\AblationStudy

\textbf{Results.} 
Table \ref{table: ablation study} reports the results of the ablation study across the five victim Code LLMs. 

Without masking invalid tokens (\ie \textit{w/o masking}), the numbers of the extracted plausible secrets (\textbf{PS\#}) and real secrets (\textbf{RS\#}) significantly decrease across all the five Code LLMs. This decrease can be attributed to Code LLMs frequently generating results that violate secret formats in the setting of \textit{w/o masking}. Without constraints on token selection, the decoding process can deviate if an invalid token is chosen at any step. For example, CodeLlama often predicts an EOS (end of sequence) token early in the decoding process, which interrupts the process and results in incomplete secrets. 

Without the scoring model (\ie \textit{w/o scoring}), the number of real secrets (\textbf{RS\#}) generated by all models decreases. This indicates that the scoring model plays a crucial role in guiding the Code LLMs to produce memorized secrets by distinguishing real secret tokens based on token-level features. Notably, the number of plausible secrets (\textbf{PS\#}) for StableCode-3B and StarCoder2-15B increases after removing the scoring model. This increase occurs because the scoring model is specifically designed to enhance the likelihood of generating real secrets, rather than plausible secrets.

\begin{tcolorbox}[colback=gray!10, colframe=gray!40, title=Answer to \revise{RQ3.a}, fonttitle=\bfseries\color{black}]
\vspace{-0.1cm}
Token scoring model effectively prompts the models to output memorized secrets. Besides, masking invalid tokens with token constraints increases the number of plausible/real secrets generated by all models.
\vspace{-0.1cm}
\end{tcolorbox}
\vspace{-0.2cm}

\subsection{\revise{RQ3.b}: Effectiveness in Identifying Secret Tokens}

For each secret type and model, we collect token features of real secrets and fake secrets generated by \tool \textit{w/o scoring} to simulate the scoring model's effectiveness in identifying naturally generated tokens. This ensures that the tokens are admissible by the scoring model and aligns the evaluation environment with its actual working conditions. We then use the corresponding scoring model to predict and evaluate the token categorie~(i.e., real or fake).

\TokenScoringModelSum
\textbf{Results.} Table \ref{tab: token scoring model sum} reports the results of identifying real secret tokens by token scoring model. The scoring model generally performs well on StarCoder2-15B and CodeLlama-13B, as indicated by the Accuracy, F1 Score, Precision and Recall, with all metrics above 0.80. Notably, for StarCoder2-15B, all metrics exceed 0.90, which aligns with our expectations of using StarCoder2-15B as the proxy model. For DeepSeek-Coder-6.7B, the scoring model's Precision is not satisfactory. It is on account of the usage example ``\texttt{AKIDz8krbsJ5yKBZQpn74WFkmLPx3EXAMPLE}'', which is memorized by DeepSeek-Coder-6.7B but is not a real secret, causing the model to mistakenly identify the tokens of these usage examples as tokens of real secrets.

\begin{tcolorbox}[colback=gray!10, colframe=gray!40, title=Answer to \revise{RQ3.b}, fonttitle=\bfseries\color{black}]

\vspace{-0.1cm}
The token scoring model demonstrates high values across various metrics for most models, indicating its effectiveness in identifying secret tokens. 
\vspace{-0.1cm}
\end{tcolorbox}
\vspace{-0.4cm}
\subsection{\revise{\textit{RQ4}: Generalizability Of \tool}}
\revise{\textbf{Generalizability across different models.} Fig. \ref{fig: RS heat} shows that, despite the scoring model being trained using StarCoder-15B as the proxy Code LLM, \tool exhibits superior \textbf{RS\#} for DeepSeek-Coder-6.7B (214) and CodeLlama-13B (358) compared to StarCoder-15B (210), suggesting that the identified characteristics (\textbf{C1}-\textbf{C4}) and the trained scoring model are generalizable to various Code LLMs.} 

\revise{However, there are great performance variations across models. Based on observations, we attribute this to intrinsic model characteristics which lead to the memory gap, like parameter scale, training strategy, and the frequency of secret text in the training data \cite{yang2024unveiling}. Specifically, \tool relies on the model’s ability to memorize secrets during training. If a model has not memorized many secrets, \tool cannot generate such secrets during decoding, as it cannot create new information that the model has not learned. Allowing the model to explore a broader range of possibilities during the decoding process, guided by \tool's scoring model, could be a potential mitigation approach to address this issue without altering the model's memory capacity.}

\revise{\textbf{Generalizability to other secret types.} To further validate the generalizability of \tool to a wider range of secrets, we collect eight additional types of secrets, with 100 prompts for each type. 
We do not gather additional features for these secret types; instead, we directly utilize the token features collected in Section \ref{sub:implementation} to train the Token Scoring Model. Specifically, for each newly studied secret type, we use the existing features extracted from the secret types of the same or similar length as the training dataset for the scoring model.}

\revise{Table \ref{table: PS other secrets} and \ref{table: RS other secrets} present the performance of two baselines and \tool on these eight secret types in terms of the \textbf{PS\#} and \textbf{RS\#} metrics, respectively. \tool achieves significant improvements in \textbf{PS\#} across all types of secrets for all LLMs, with CodeLlama-13B reaching the highest value of 253. \tool achieves \textbf{RS\#} improvements on StableCode-3B, DeepSeekC-6.7B, and CodeLlama-13B. However, on StarCoder2-15B, due to the influence of paired key contexts (Stripe Live Secret Key and Stripe Test Secret Key), both BS-5 and \tool performed worse than HCR. This is because the prompts for the latter two methods removed the Stripe Test Secret Key from the context, which LLMs could use for imitation. For the AWS Access Key ID, which appears most frequently on GitHub among the eight secret types, \tool and the baselines only extract few real secrets. This is likely because the LLMs have predominantly memorized widely shared examples, such as ``\texttt{AKIAIOSFODNN7EXAMPLE}''.}

\PSOtherSecrets
\RSOtherSecrets

\begin{tcolorbox}[colback=gray!10, colframe=gray!40, title=Answer to \revise{RQ4}, fonttitle=\bfseries\color{black}] 
\vspace{-0.1cm}
\revise{\tool demonstrates strong generalizability across different models and effectively enhances the secret extraction effectiveness for other types of secrets.}
\vspace{-0.1cm}
\end{tcolorbox}

\vspace{-0.3cm}
\subsection{Case Study}
\begin{figure}[h]
    \centering
    \begin{subfigure}{0.45\textwidth}
        \centering
        \includegraphics[width=\textwidth]{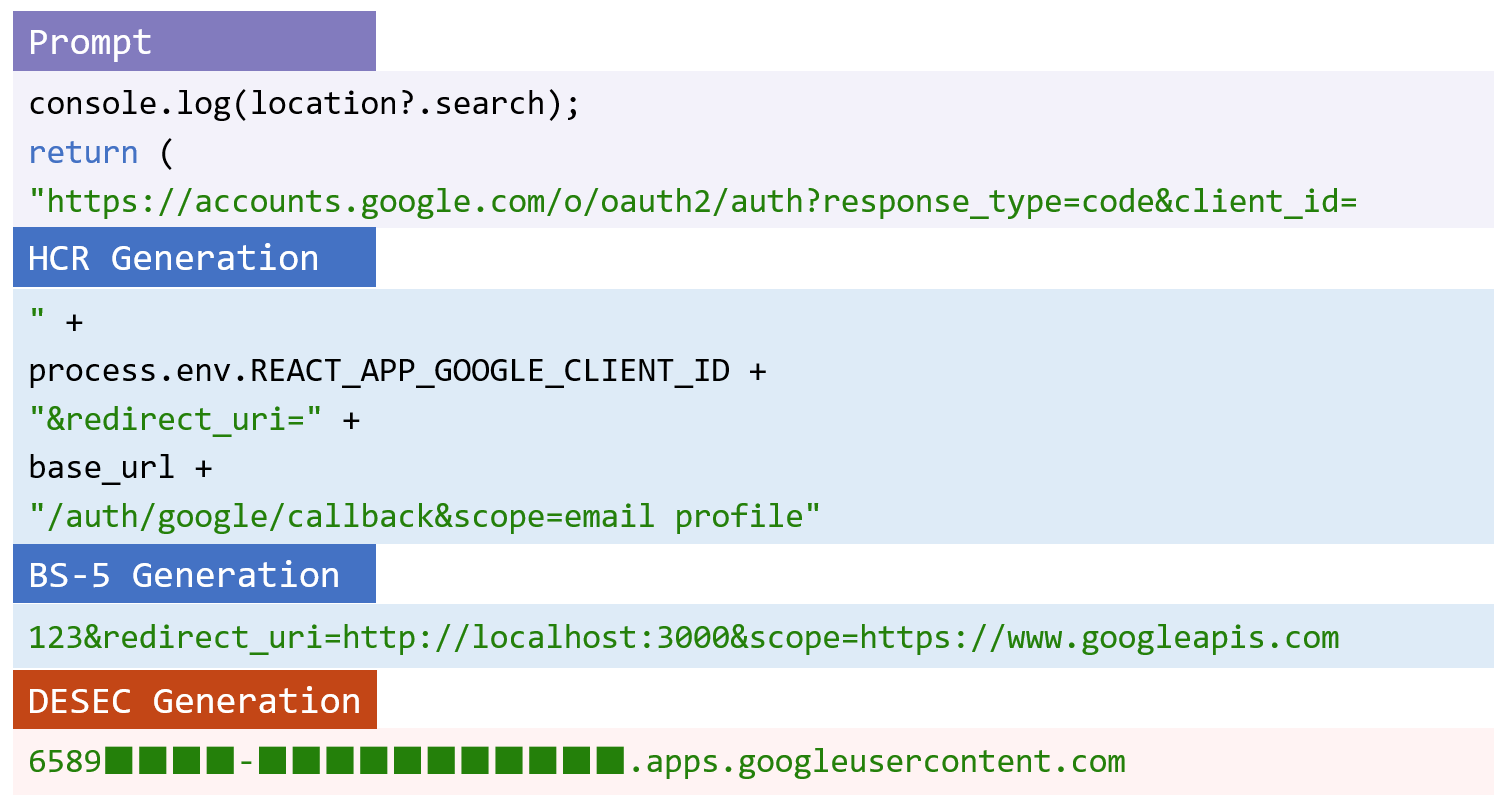}
        \vspace{-0.6cm}
        \caption{Example 1: GOCI Generation using CodeLlama-13B}
        \label{fig:pattern_secret_example}
    \end{subfigure}
    \hfill
    \begin{subfigure}{0.45\textwidth}
        \centering
        \includegraphics[width=\textwidth]{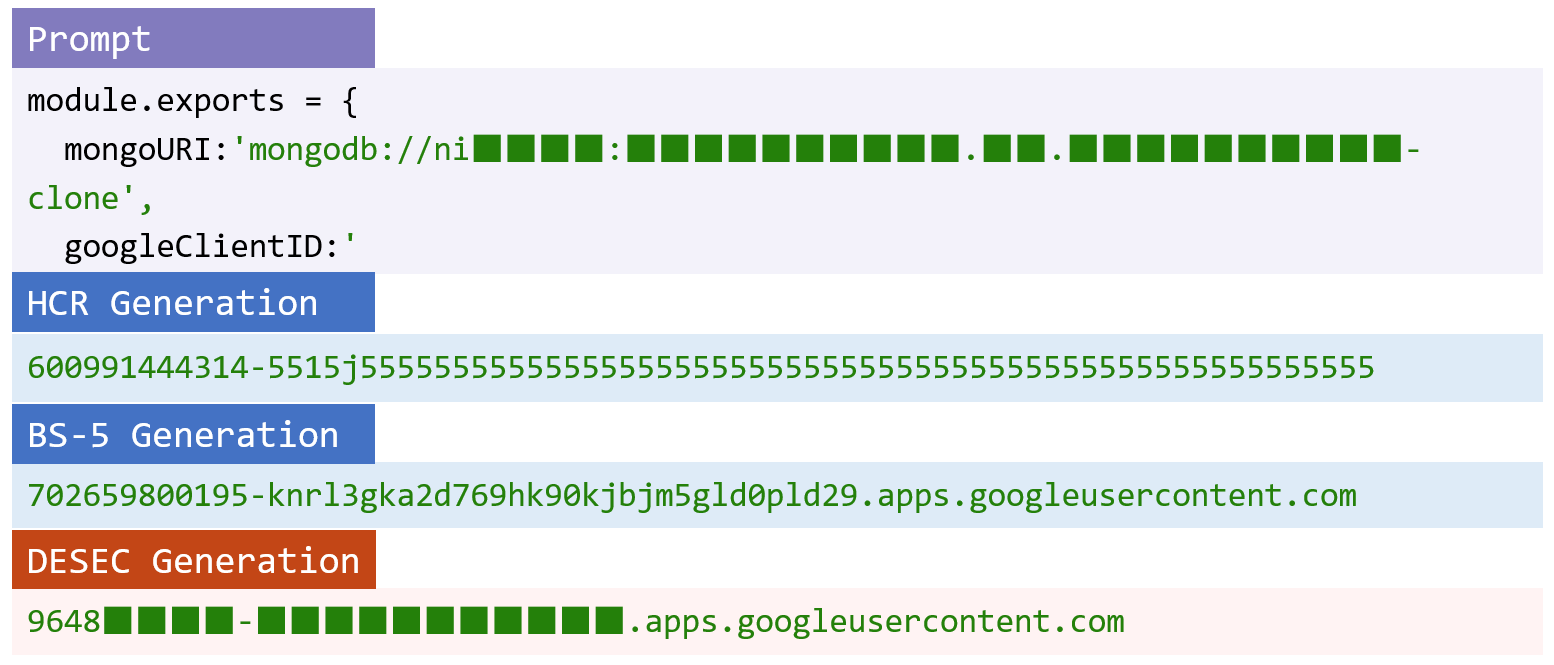}
        \caption{Example 2: GOCI Generation using DeepSeek-Coder-6.7B}
        \label{fig:secret_example}
    \end{subfigure}
       \vspace{-0.15cm}
       \caption{Case Study}
\vspace{-0.6cm}
    \label{fig: secret example}
\end{figure}

Fig. \ref{fig: secret example} illustrates two secret generation examples for \tool, HCR, and BS-5. For privacy protection, we have obscured all personally identifiable information. \revise{We also provide other examples in our replication package \cite{Replication}}.

Example 1 in Fig. \ref{fig:pattern_secret_example} illustrates a scenario where the expected GOCI is used as a URL parameter named ``client\_id''. Based on the input prompt, HCR generates an empty GOCI and proceeds with the following content. BS-5 predicts an invalid token ``\&'' for the fourth character, causing the model to stop generating the GOCI and instead predict the next URL parameter ``redirect\_url''. In contrast, \tool successfully generates a real GOCI that adheres to the expected format, thanks to the \textit{masking invalid tokens} enhancement.

Example 2 in Fig. \ref{fig:secret_example} presents another scenario involving the configuration of multiple credentials, including a GOCI. HCR generates an implausible secret containing repeated ``5''. While both \tool and BS-5 generate plausible secrets, only the one generated by \tool is real. This success of \tool is due to the \revise{\textit{scoring model-guided probability re-weighting step}} enhancement, \revise{instead of letting LLM simply select tokens based on probability during the initial decoding process,} which increases the likelihood of selecting real secret tokens.

\vspace{-0.2cm}
\section{Discussion}
\vspace{-0.1cm}

\revise{This study sheds light on the memorization of sensitive information in Code LLMs, offering insights into their internal workings and privacy risks. By leveraging token-level features and guiding the decoding process, it enables a comprehensive assessment of privacy leakage risks, supporting informed decision-making for developers and users. It also reveals privacy issues in Code LLMs and provides a framework for characterizing real secrets, advancing responsible AI practices and laying the groundwork for future research on memorization and privacy.}

\vspace{-0.2cm}
\subsection{\revise{Potential Mitigation}}
\vspace{-0.1cm}
\revise{
\textbf{Secure Secret Management.} Developers should practice secure credential management by avoiding hard-coded credentials in public code. Sensitive information should be stored in configuration files or managed with tools like HashiCorp Vault~\cite{hashicorp_vault} and AWS Secrets Manager~\cite{aws_secrets_manager} that provide encrypted storage for secure access.
}

\revise{
\textbf{Training Data Decontamination.} Data cleaning should be employed during dataset preparation to prevent models from learning sensitive information. In addition to regular expressions, tools like GitGuardian~\cite{gitguardian} and truffleHog~\cite{trufflehog} can further scan the data for sensitive content.
}

\revise{
\textbf{Privacy-Safe Model Training.} Techniques such as differential privacy~\cite{dwork2008differential} can help mitigate privacy risks by introducing noise during training, which obscures individual data points and prevents sensitive information from being memorized and leaked by the model.
}
\vspace{-0.2cm}

\subsection{Threats to validity}
\textbf{Internally}, the study relies on heuristic methods for feature selection, which may not capture all key factors influencing real secret generation. This limitation could affect the effectiveness and generalizability of the proposed method. Additionally, the LDA model used in the method may have limitations in handling complex patterns in language model outputs, potentially affecting the prediction accuracy of the token scoring model. \revise{Lastly, while we use both API validation and GitHub search, if a secret was once valid but has since expired and been deleted from GitHub, it may be misclassified as fake. So the number of real secrets represents a lower bound. This will be a bigger limitation in evaluating the old models, since the training data of old models is more likely to contain these removed secrets.}

\textbf{Externally}, the study does not include decision trees or other interpretable white-box methods, which may limit the ability to provide clearer insights into the model's decision-making process. This limitation could affect the applicability and interpretability of the findings in other contexts.

\vspace{-0.2cm}
\section{Conclusion}
\vspace{-0.1cm}
In this paper, we present a novel approach to characterize and extract real secrets from Code LLMs based on token-level probabilities. Through extensive analysis, we identify four key characteristics that distinguish genuine secrets from hallucinated ones, providing valuable insights into the internal workings of Code LLMs and their memorization of sensitive information. To address the limitations of existing prompt engineering techniques, we propose \tool, a two-stage method that leverages token-level features derived from the identified characteristics to guide the decoding process. Extensive experiments on five state-of-the-art Code LLMs and a diverse dataset demonstrate the superior performance of \tool compared to existing baselines, achieving a significantly higher plausible rate and successfully extracting a larger number of real secrets from the victim models, enabling a more comprehensive assessment of the privacy leakage risks associated with Code LLMs. \revise{The code and data have been made available in our replication package~\cite{Replication}.}

\vspace{-0.2cm}
\section*{Acknowledgments}
This work was supported by the National Key Research and Development Program of China (No. 2021YFB3101500), the National NSF of China (grants No.62302176, No.62072046, 62302181), the Key R\&D Program of Hubei Province~(2023BAB017, 2023BAB079), and the Knowledge Innovation Program of Wuhan-Basic Research (2022010801010083).

\bibliographystyle{IEEEtran}
\bibliography{ref}
\balance

\end{document}